\documentclass[useAMS,usenatbib,usegraphicx]{mn2e} 
\usepackage{amssymb} 
\usepackage{times} 
\usepackage{aas_macros} 
\usepackage[usenames]{color} 

\title[KIC\,10195926]{The first evidence for multiple pulsation axes: a new roAp 
star in the {\it Kepler} field, KIC\,10195926}

\author[D.W. Kurtz et al.] 
{D.W. Kurtz$^{1}$, M.S. Cunha$^{2}$, H. Saio$^{3}$, L. Bigot$^{4}$, L. A. 
Balona$^{5}$, V.G. Elkin$^{1}$, H. Shibahashi$^{6}$,  \\
\newauthor{ I.M. Brand\~ao$^{2}$, K. Uytterhoeven$^{7}$, S. 
Frandsen$^{8}$, S. Frimann$^{8,9}$, A. Hatzes$^{10}$,T. Lueftinger$^{11}$,}\\
\newauthor{ M. Gruberbauer$^{12}$}, H. Kjeldsen$^{8}$, 
J. Christensen-Dalsgaard$^{8}$, S.D. Kawaler$^{13}$ \\ 
\\
\\
$^{1}$Jeremiah Horrocks Institute of Astrophysics, University of Central 
Lancashire, Preston PR1\,2HE, UK\\ 
$^{2}$Centro de Astrof\'\i sica e Faculdade de Ci\^encias, Universidade do Porto, 
Rua das Estrelas, 4150-762, Portugal \\ 
$^{3}$Astronomical Institute, Graduate School of Science, Tohoku University, 
Sendai, 980-8578, Japan \\ 
$^{4}$Universit\'e Nice Sophia-Antipolis, CNRS UMR 6202, Observatoire de la
C\^ote d'Azur, BP 4229, 06304 Nice, France \\
$^{5}$South African Astronomical Observatory, P.O. Box 9, Observatory 7935, Cape 
Town, South Africa \\ 
$^{6}$Department of Astronomy, {University of Tokyo}, Tokyo 113-0033, Japan\\
$^{7}$Laboratoire AIM, CEA/DSM-CNRS-Universit\'e Paris Diderot; CEA, IRFU, SAp, 
centre de Saclay, F-91191, Gif-sur-Yvette, France \\ 
$^{9}$Nordic Optical Telescope, 38700 Santa Cruz de la Palma, Spain \\
$^{8}$Department of Physics and Astronomy, Ny Munkegade 120, Aarhus University, 
8000 Aarhus C, Denmark \\ 
$^{10}$Th\"uringer Landessternwarte Tautenburg, Sternwarte 5, 07778 Tautenburg, 
Germany \\ 
$^{11}$Astronomisches Institut der Universit\"at Wien, T\"urkenschanzstr. 17, 
A-1180 Wien, Austria \\ 
$^{12}$Department of Astronomy \& Physics, Saint Mary's University, Halifax, 
NS\,B3H\,3C3 Canada \\ 
$^{13}$Department of Physics and Astronomy, Iowa State University, Ames, IA 5011, 
USA \\
} 
 
\begin{document} 
 
\maketitle 

\begin{abstract} 
We have discovered a new rapidly oscillating Ap star among the {\it Kepler} 
Mission target stars, KIC\,10195926. This star shows two pulsation modes with 
periods that are amongst the longest known for roAp stars at 17.1\,min and 
18.1\,min, indicating that the star is near the terminal age main sequence. The 
principal pulsation mode is an oblique dipole mode that shows a rotationally split 
frequency septuplet that provides information on the geometry of the mode. The 
secondary mode also appears to be a dipole mode with a rotationally split triplet, 
but we are able to show within the improved oblique pulsator model that these two 
modes cannot have the same axis of pulsation. This is the first time for any 
pulsating star that evidence has been found for separate pulsation axes for 
different modes. The two modes are separated in frequency by 55\,$\mu$Hz, which we 
model as the large separation. The star is an $\alpha^2$\,CVn spotted magnetic 
variable that shows a complex rotational light variation with a period of $P_{\rm 
rot} = 5.68459$\,d. For the first time for any spotted magnetic star of the upper 
main sequence, we find clear evidence of light variation with a period of twice 
the rotation period; i.e. a subharmonic frequency of $\nu_{\rm rot}/2$. We propose 
that this and other subharmonics are the first observed manifestation of torsional 
modes in an roAp star. From high resolution spectra we determine $T_{\rm eff} = 
7400$\,K, $\log g = 3.6$ and $v \sin i = 21$\,km\,s$^{-1}$. We have found a 
magnetic pulsation model with fundamental parameters close to these values that 
reproduces the rotational variations of the two obliquely pulsating modes with 
different pulsation axes. The star shows overabundances of the rare earth 
elements, but these are not as extreme as most other roAp stars. The spectrum is 
variable with rotation, indicating surface abundance patches. 
\end{abstract} 
 
 
\section{Introduction} 
 
The rapidly oscillating Ap (roAp) stars are a unique laboratory in which the 
interaction between stellar pulsation and a magnetic field can be studied in 
greater detail than in any other star except the Sun. Although the pulsation 
periods in roAp stars are about the same as in the Sun (a few minutes), they are 
probably driven by the $\kappa$-mechanism in the hydrogen ionization zone, rather 
than being stochastically excited by convection, as in the Sun. It is thought that 
suppression of convection at the magnetic poles reduces damping in the hydrogen 
ionization zone and allows modes of high radial order to be driven 
\citep{Balmforth2001}. 

{However, when comparing theoretical instability strips to the positions of the 
known roAp stars in the HR diagram,  there remains a difficulty in explaining the 
presence of pulsations in the coolest roAp stars. The theoretical red edge seems 
to be hotter than its observed counterpart, a fact that was noted when the first 
theoretical instability strip was computed \citep{cunha02}, and that is even more 
obvious today, with more roAp stars having reliable effective temperature 
determinations below 7000\,K. Moreover, the agreement does not seem to improve 
when models with different global metallicity or local metal accumulation are 
considered \citep{theadoetal09}. While the iron abundance may influence the 
excitation, it seems not to have an impact on the red edge of the theoretical 
instability strip.}
 
The Ap stars have strong, global magnetic fields in the range $1 - 25$\,kG and 
overabundances of some rare earth elements that can exceed $10^6$ of the solar 
value. These abundance anomalies are located in patches on the stellar surface 
that give rise to photometrically observable rotational light variations. The 
abundances are also vertically stratified in the observable atmosphere. They are 
thought to arise by atomic diffusion in the presence of a magnetic field. Thus Ap 
stars in general, and roAp stars in particular, are test beds for atomic diffusion 
theory, which is important for solar physics, stellar cluster age determinations, 
pulsation driving in main sequence and subdwarf B stars and a possible solution 
for frequency anomalies in $\beta$\,Cep stars. 
 
The general view of roAp stars is that the pulsation axis is not aligned with the 
rotation axis, thus uniquely allowing the pulsation modes to be viewed from 
varying aspect with rotation. This provides geometric mode identification 
information that is not available for other asteroseismic targets. A simple model 
is that the pulsation axis is aligned with the axis of the magnetic field, which 
is assumed to be roughly a dipole inclined with respect to the axis of rotation. 
As the star rotates, the geometry of the pulsation changes, leading to amplitude 
and, in some cases, phase modulation. This is the oblique pulsator model 
\citep{kurtz82}. 

In early work it was assumed that the eigenfunction could be represented as a pure 
axisymmetric spherical harmonic, in which case the variation of pulsational 
amplitude with rotation leads directly to a determination of the spherical 
harmonic degree, $l$. Furthermore, amplitude variation also allows constraints on 
the relative inclinations of the rotational and magnetic axes to be placed. We now 
know that the eigenfunction is far from a simple spherical harmonic 
\citep{takatashiba94, takatashiba95, dziembowski96, cunha06, 
saiogautschy04, bigotetal00} 
and values derived in this way should be treated with caution. Treating the 
combined effects of rotation and magnetic field on the pulsation modes is complex. 
\citet{bigot02} showed that in the case of a relative weak global magnetic field 
-- $B_{\rm p} \le 1000$\,G -- the pulsation axis is, in general, neither the 
magnetic axis nor the rotation axis, but lies between these. 
 
There are 40 roAp stars known from ground-based observations that are listed by 
\citet{kurtzetal06}, and a few discovered more recently (see, e.g., 
\citealt{kochukhovetal09}; \citealt{elkinetal10}). High-resolution spectroscopy 
using large telescopes allows sufficient time resolution for the detailed study of 
line profile variations caused by rapid pulsation. It turns out that detection of 
low-amplitude roAp pulsations is easier using radial velocities obtained in this 
way rather than high-precision photometry. Several roAp stars are known in which 
the radial velocity variations can be detected, but no corresponding light 
variations can be seen from ground-based observations. A global network of 
spectrographs dedicated to asteroseismology, such as the SONG (Stellar 
Observations Network Group) project (e.g., \citealt{grundahletal09}) may greatly 
increase the number of known roAp stars. For now, however, photometry is still the 
most cost-effective method for global asteroseismic studies in these stars where 
the frequencies are the fundamental data. For detailed examination of pulsation 
mode behaviour in the observable atmospheres, spectroscopic studies are superior 
(see, e.g., \citealt{ryabchikovaetal07}, \citealt{freyhammeretal09}). 
 
\begin{figure*} 
\centering 
\includegraphics[width=16cm]{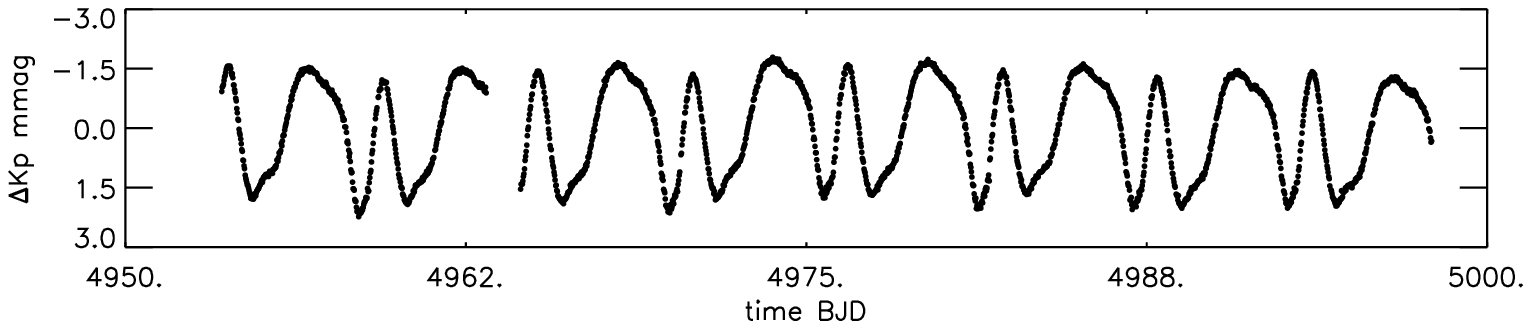} 
\includegraphics[width=16cm]{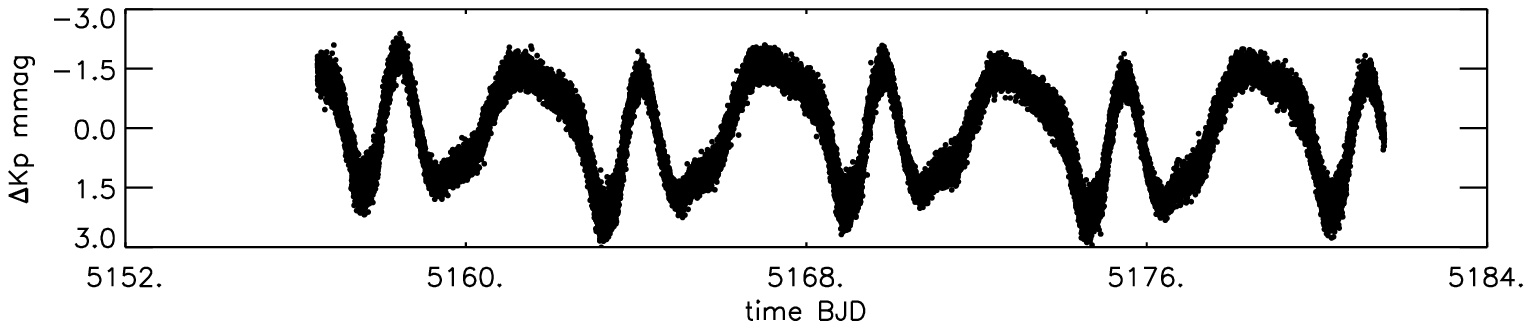} 
\includegraphics[width=16cm]{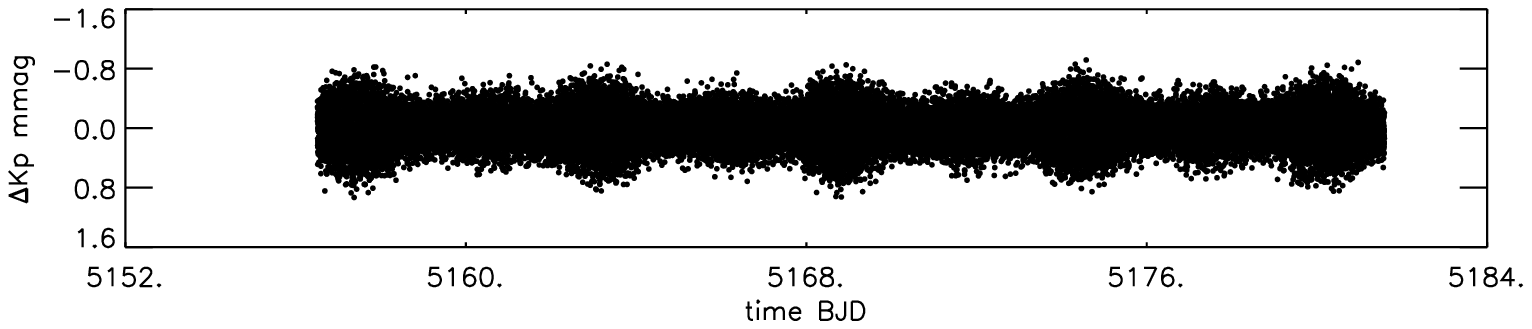} 
\caption{Top: The long cadence light curve of KIC\,10195926 from Q0 + Q1. The 
ordinate is ${\rm BJD} - 2450000$. Middle: The short cadence light curve of 
KIC\,10195926 from Q3.3. Bottom: the short cadence light curve after high pass 
filtering (described in Section\,\ref{sec:pulsations}) has removed the rotational 
light variations, leaving only the pulsational variations. Only the envelope of 
the pulsation is visible here, as the pulsation period is too short to be resolved 
at this scale. Note that pulsation maximum coincides with rotational minimum 
light. } 
\label{fig:lc} 
\end{figure*} 
 
The {\it Kepler} Mission has revolutionised our ability to detect low-amplitude 
light variations in stars, reducing the detection limit to only a few 
micromagnitudes even in rather faint stars. The {\it Kepler} asteroseismic 
programme is described by \citet{gillilandetal2010}, and details of the mission 
can be found in \citet{kochetal2010}. {\it Kepler} is observing about 150\,000 
stars continuously over its minimum 3.5-y mission lifetime with 30-min time 
resolution. A small allocation (512 stars) can be studied with 1-min time 
resolution. While there were no known roAp stars in the {\it Kepler} field before 
launch in 2009, three have been found in the 1-min cadence data. KIC\,8677585 was 
studied by \citet{balonaetal10} who found several pulsation frequencies with 
periods around 10\,min, including an unexpected low frequency of 3.142\,d$^{-1}$ 
which is possibly a g-mode pulsation. A second roAp star discovered in the {\it 
Kepler} data, KIC\,10483436, shows a rotational light variation and two frequency 
multiplets (Balona et al., in preparation). 
 
The third roAp star discovered in the {\it Kepler} data is KIC\,10195926 ($\alpha 
= 19\,05\,27$, $\delta = +47\,15\,48$, J2000; $Kp = 10.57$), which we report on 
here. It has a principal mode that shows a clear frequency septuplet with 
separations equal to the rotation period of the star, $P_{\rm rot} = 5.68459$\,d. 
Rotational light variations are clearly seen, indicating that it is an 
$\alpha^2$\,CVn star. It appears that in this star both pulsation poles can be 
viewed over the rotation cycle, which resembles the well-studied roAp star 
HR\,3831 (see, e.g., \citealt{kurtzetal90}, \citealt{kurtzetal97}, 
\citealt{kochukhovetal04}, \citealt{kochukhov06}, and references therein). A 
second, much lower amplitude mode that shows a frequency triplet split by the 
rotation frequency is also found, thanks to the 5\,$\mu$mag precision of the {\it 
Kepler} data. Interestingly, it appears that these two modes do not share the same 
pulsation axis, an effect that has never been detected previously. We present 
theoretical support for this hypothesis. 

\section{Fundamental parameters from spectroscopy} 
\label{sec:spectroscopy} 

Spectroscopic observations of KIC\,10195926 were made with the high-resolution 
Fibre-fed Echelle Spectrograph (FIES) at the Nordic Optical Telescope (NOT) in 
2010 June, July and August. The spectra have a resolution of $R = 65\,000$ and 
cover the spectral range $370 - 730$\,nm. The signal-to-noise ratio (S/N) attained 
is about 60 for the June spectra and 40 for observations in July and August. The 
star was also observed with the Sandiford Echelle spectrograph on the 2.1-m 
telescope at McDonald Observatory. The spectra have a resolution of $R = 60\,000$ 
and were collected on four nights in 2010 July covering the spectral region $508 - 
605$\,nm. On nights when multiple exposures were taken, they were co-added. The 
S/N for these observations ranges from 50 to 80. Table\,\ref{obslog} gives a 
journal of the observations indicating their rotational phase (see 
Section\,\ref{sec:lightcurves} below).

\begin{table} 
\caption[]{A journal of spectroscopic observations obtained with FIES on the 
Nordic Optical Telescope, and with the Sandiford Spectrograph on the McDonald 
Observatory 2.1-m telescope. The rotational phase has been calculated using the 
ephemeris $\Phi_{\rm rot} = 245\,5168.91183 + 5.68459$. } 
\begin{center} 
\begin{tabular}{ccc} 
\hline 
\multicolumn{3}{c}{Nordic Optical Telescope -- FIES}\\
\hline 
\multicolumn{1}{c}{JD $- 2450000.0$} & 
\multicolumn{1}{c}{exposure time (s)} & 
\multicolumn{1}{c}{rotational phase}\\
5369.64536 & 1800 & 0.31 \\ 
5369.66862 & 1800 & 0.32 \\ 
5407.37706 & 1800 & 0.95 \\ 
5407.40009 & 1800 & 0.95 \\ 
5408.37815 & 1800 & 0.13 \\ 
5408.40118 & 1800 & 0.13 \\ 
5411.68126 & 1800 & 0.71 \\ 
5411.70415 & 1800 & 0.71 \\ 
\hline 
\multicolumn{3}{c}{McDonald 2.1-m telescope -- Sandiford spectrograph}\\
\hline 
5397.92206 & 2400 & 0.28 \\ 
5398.69240 & 1800 & 0.42 \\ 
5399.70613 & 1800 & 0.60 \\ 
5400.90493 & 1800 & 0.81 \\ 
\hline 
\end{tabular} 
\label{obslog} 
\end{center} 
\end{table} 

Most roAp stars have strong magnetic fields in the range of $1 - 25$\,kG. 
Knowledge of the magnetic field strength is important for understanding and 
modelling these stars. In the spectra of KIC\,10195926 there is no obvious Zeeman 
splitting of magnetically sensitive spectral lines. We therefore searched for 
magnetic broadening by comparing the observations with synthetic spectra 
calculated for the spectral region around the Fe\,\textsc{ii} 6149.258\,\AA\ line 
using the {\small SYNTHMAG} program by \citet{piskunov99}. With the S/N of our 
spectra and rotational broadening, we can only place an upper limit on the 
magnetic field modulus of about 5\,kG. 
 
Because of the probable magnetic broadening, it is necessary to use lines with 
negligible Land\'e factors to measure the projected rotational velocity $v \sin 
i$. Because of the patchy distribution of elements over the surface, it is 
necessary to examine spectra at different rotation phases, since lines from 
elements concentrated in spots do not always sample the full rotation velocity of 
the star. From the Fe\,\textsc{i} line at 5434.523\,\AA\ we determined from 
spectra at different rotation phases that $v \sin i = 21 - 24$\,km\,s$^{-1}$. Most 
spectral lines are fitted well with values in this range, but some require an even 
smaller $v \sin i$, probably because of the spots. At some rotation phases 
24\,km\,s$^{-1}$ is needed to fit the line broadening. Our internal error on a 
determination of $v \sin i$ is 1\,km\,s$^{-1}$, which we cautiously increase to 
2\,km\,s$^{-1}$. Our estimate is $v \sin i = 21 \pm 2$\,km\,s$^{-1}$, but we 
consider this to be near a lower limit for the equatorial rotation velocity, given 
the larger $v \sin i$ that we see at some rotation phases. 

KIC\,10195926 has photometrically determined parameters from broadband photometry 
given in the Kepler Input Catalogue (KIC): $Kp = 10.57$, $T_{\rm eff} = 7400$\,K, 
$\log g = 3.6$. The KIC gives an estimate of the contamination factor for this 
star of 0.015, meaning that less than 1.5 percent of the light in the observing 
mask can be from contaminating background stars. This makes it exceedingly 
improbable that any of the light variations we discuss in the paper might be from 
a contaminating background star.

To determine the effective temperature of the star, Balmer lines profiles were 
compared with synthetic profiles starting with parameters close to those from 
Kepler Input Catalogue. Synthetic spectra were calculated with the {\small SYNTH} 
program of \citet{piskunov92}. We used model atmospheres from the NEMO database 
(Vienna New Model Grid of Stellar Atmospheres) \citep{heiteretal02}. This grid has 
an effective temperature step of 200\,K, which we interpolated to get models with 
a 100\,K step. The spectral line list for analysis and {\small SYNTH} calculations 
was taken from the Vienna Atomic Line Database (VALD, \citealt{kupkaetal99}), 
which includes lines of rare earth elements from the DREAM database 
\citep{biemontetal99}. 
 
We found a best fit for H$\alpha$ and H$\beta$ with synthetic profiles for $T_{\rm 
eff} = 7200$\,K, with estimated error of 200\,K. The determination of the surface 
gravity is difficult at this temperature. The Balmer line profiles are relatively 
insensitive to $\log g$, constraining it only to within $\pm 0.5$. Ionisation 
equilibrium for Fe\,\textsc{i} and Fe\,\textsc{ii} lines and Cr\,\textsc{i} and 
Cr\,\textsc{ii} lines is potentially much more sensitive for determining $\log g$, 
but we did not get reliable results with our initial study. We expect a more 
detailed analysis, which is underway, will better constrain $\log g$. We therefore 
use here the photometrically determined value of $\log g = 3.6$ (cgs) from the 
KIC. Because the spectrum is not very peculiar, we estimate a 1$\sigma$ error of 
$\pm 0.3$.

A detailed abundance analysis is in progress. We made preliminary estimates of the 
abundances of some elements at two rotation phases to get an overview of the 
spectral character of KIC\,10195926. Using synthetic spectra calculated with 
{\small SYNTH} for model $T_{\rm eff} = 7200$\,K and $\log g = 3.6$ we examined 
some abundances using the first and third NOT spectra listed in 
Table\,\ref{obslog}. The spectral lines are clearly variable with rotation, so 
other spectra will give values that differ for some elements. Surprisingly for a 
roAp star, there are no obvious lines of rare earth elements. The usual strong 
lines in roAp stars of Nd\,\textsc{iii} and Eu\,\textsc{ii} are present and 
indicate overabundances compared to the Sun, but less than is typical for most 
other roAp stars. Si\,\textsc{ii} lines are normal. There is no obvious core-wing 
anomaly \citep{cowleyetal01} in H$\alpha$, a usual signature of roAp stars. 

\begin{figure} 
\centering 
\includegraphics[width=8cm]{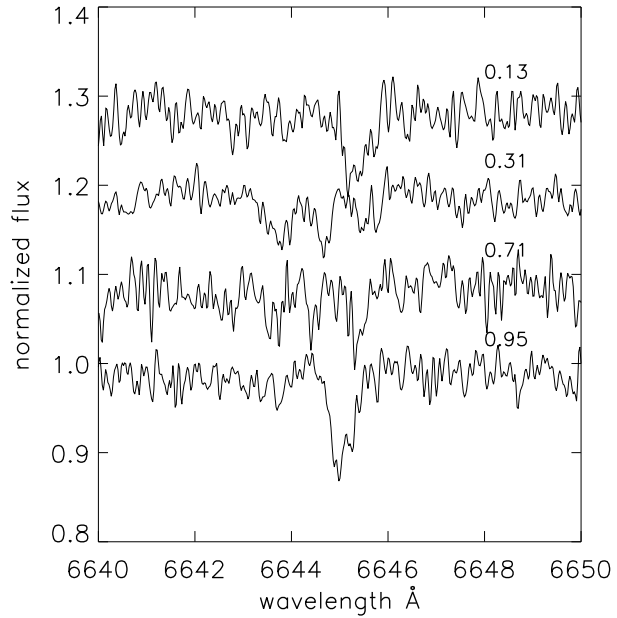} 
\caption{This shows the Eu\,\textsc{ii} 6645\,\AA\ line at four different rotation 
phases. The spectra have been offset on the y-axis for visibility. The bottom 
spectrum is for rotation phase 0.95, which is close to the time of pulsation 
maximum and also rotational light minimum (see Fig.\,\ref{fig:phase-q01}.) This is 
when one of the Eu\,\textsc{ii} spots is close to crossing the line-of-sight. In 
the middle panels, which are near quadrature, it is likely that two small spots of 
Eu\,\textsc{ii} that are concentrated near the magnetic poles are seen separately. 
This will be studied in detail with future, higher S/N spectra.} 
\label{fig:sp6645} 
\end{figure}

Table\,\ref{abund} gives the abundances we found compared to solar values. The 
second column gives the abundances for a spectrum obtained at rotational phase 
0.31. For this spectrum no clear evidence lines of rare earth elements was found. 
There are some hints of the presence of such lines in the spectrum, but the better 
S/N spectra are required. The fourth column gives abundances for a spectrum 
obtained at rotational phase 0.95. This spectrum definitely shows lines of 
Eu\,\textsc{ii} and a trace of weak lines of Nd\,\textsc{iii} which confirm the 
peculiarity of this star. Lines of strontium are much wider and stronger in this 
spectrum while the barium lines are much weaker when compared with the spectrum at 
phase 0.31. Iron lines are weaker and slightly wider with lower abundances. The 
spectral lines show strong variability with rotational period, demonstrating the 
spotted surface structure. By comparing with the roAp star HD\,99563 
\citep{freyhammeretal09}, we judge that lines of Eu\,\textsc{ii} and 
Ba\,\textsc{ii} are formed in rather small spots. Other elements also demonstrate 
nonuniform distribution over stellar surface. Fig.\,\ref{fig:sp6645} shows the 
Eu\,\textsc{ii} 6645-\AA\ line at four rotation phases. While better S/N is needed 
for further study, it is clear that Eu\,\textsc{ii} is concentrated is at least 
two small spots close to the pulsation pole (and hence probably close to the 
magnetic pole) of this star.

{We note that the abundances determined here are not characteristic of the entire 
star. Atomic diffusion causes atmospheric stratification in Ap stars, with some 
elements being enhanced by orders of magnitude with respect to the Sun, while 
others are deficient. As a main sequence A star KIC\,10195926 has an age less than 
1\,Gyr, hence is expected to have global abundances somewhat higher than those of 
the much-older Sun. This is a consideration when modelling the star.}

\begin{table} 
\caption[]{Chemical abundances for KIC\,10195926 and the corresponding solar 
abundances \citep{asplundetal09}. The errors quoted are internal standard 
deviations for the set of lines measured. Columns 2 and 4 give, respectively, 
abundances determined at two rotation phases, 0.31 and 0.95, using $P_{\rm rot} = 
5.68459$\,d (see section\,\ref{sec:lightcurves}) . The rotation phase refers to 
the time of pulsation maximum, which coincides with rotational light minimum as 
seen in Fig.\,\ref{fig:phase-q01}. Columns 3 and 5 give the number of lines, ${\rm 
N}_{\rm l}$, used in each case.} 
\begin{center}  
\begin{tabular}{lcrrrr}  
\hline   
\multicolumn{1}{c}{Ion} &  
\multicolumn{1}{c}{$\log\,N/N_{\rm tot}$} & 
\multicolumn{1}{c}{${\rm N}_{\rm l}$} &  
\multicolumn{1}{c}{$\log\,N/N_{\rm tot}$} &  
\multicolumn{1}{c}{${\rm N}_{\rm l}$} &  
\multicolumn{1}{c}{$\log\,N/N_{\rm tot}$} \\  
& \multicolumn{1}{c}{$\phi=0.31$}  & 
& \multicolumn{1}{c}{$\phi=0.95$}  &
& \multicolumn{1}{c}{Sun} \\   
\hline 
C\,\textsc{i}  & $-3.6 \pm 0.2 $ & 2 &  &  & $-3.57$ \\  
Mg\,\textsc{i}  & $-4.4 \pm 0.1 $ & 3 & $ -4.3 \pm 0.1 $ & 4 & $-4.47$ \\  
Si\,\textsc{i}  & $-4.4 \pm 0.2 $ & 3 & $ -4.1 \pm 0.1 $ & 4 & $-4.40$ \\ 
Si\,\textsc{ii} & $-4.2 \pm 0.1 $ & 3 & $ -4.0 \pm 0.2 $ & 3 & $-4.49$ \\ 
Ca\,\textsc{i}  & $-5.1 \pm 0.1 $ & 18 & $ -5.3 \pm 0.2 $ & 10 & $-5.66$ \\ 
Sc\,\textsc{ii} & $-8.7 \pm 0.1 $ & 3 & $ -9.9 \pm 0.2 $ & 2 & $-8.85$ \\ 
Ti\,\textsc{ii} & $-6.8 \pm 0.3 $ & 3 &   &  & $-7.05$ \\ 
Cr\,\textsc{i} & $-5.8 \pm 0.2 $ & 4 & $ -5.4 \pm 0.4 $ & 4 & $-6.36$ \\ 
Cr\,\textsc{ii} & $-5.7 \pm 0.1 $ & 6 & $ -5.7 \pm 0.1 $ & 8 & $-6.36$ \\ 
Fe\,\textsc{i}  & $-4.3 \pm 0.1 $ & 25 & $ -4.6 \pm 0.1 $ & 29 & $-4.50$ \\ 
Fe\,\textsc{ii} & $-4.3 \pm 0.1 $ & 7 & $ -4.6 \pm 0.1 $ & 7 & $-4.50$ \\ 
Ni\,\textsc{i}  & $-5.9 \pm 0.1 $ & 4 & $ -6.7 \pm 0.2 $ & 3 & $-5.78$ \\ 
Sr\,\textsc{ii} & $-8.8 \pm 0.2 $ & 2 & $ -8.0 \pm 0.2 $ & 2 & $-9.13$ \\ 
Y\,\textsc{ii}  & $-8.9 \pm 0.1 $ & 2 & $ -8.8 \pm 0.3 $ & 5 & $-9.79$ \\ 
Ba\,\textsc{ii} & $-9.0 \pm 0.1 $ & 3 & $ -10.4 \pm 0.3 $ & 2 & $-9.82$ \\
Nd\,\textsc{iii} & $       $ &   & $ -9.3 \pm 0.1 $ & 3 & $-10.58$ \\
Eu\,\textsc{ii} & $       $ &   & $ -8.4 \pm 0.1 $ & 4 & $-11.48$ \\
\hline   
\end{tabular}   
\label{abund}  
\end{center} 
\end{table} 

\section{The light curves and rotation period} 
\label{sec:lightcurves} 
 
KIC\,10195926 has been observed by the {\it Kepler} mission during three ``rolls'' 
up to this writing. Following each quarter of a 370-d {\it Kepler} solar orbit the 
telescope is rolled to keep its solar panels facing the Sun. Choices and changes 
of targets are made on the basis of these ``quarters'', which sometimes are split 
into approximately 1-month ``thirds''. Most data (for $>$150\,000 stars) are 
obtained in long cadence (LC) with integration times of 29.4\,min. For 512 targets 
data are obtained in short cadence (SC) mode with sampling times just under 1-min. 
The three rolls for which we have data for KIC\,10195926 are long cadence in Q0 
(quarter zero), which was a commissioning roll of less than 10\,d, long cadence in 
Q1, and short cadence in Q3.3 (25\,d during the last third of Q3). 
Table\,\ref{tab:journal} gives a journal of the data. 
 
\begin{table} 
\caption[]{A journal of the {\it Kepler} data for KIC\,10195926. The third column 
is the number of data points after a few obvious outliers have been removed. The 
Q1 data have had a linear trend removed because of instrumental drift. The final 
column gives the root-mean-square variance in each data set of one point with 
respect to the mean. This includes all variance in noise and stellar variability.} 
\label{tab:journal}  
\begin{center} 
\begin{tabular}{ccrrcc} 
\hline 
\multicolumn{1}{c}{quarter} 
& \multicolumn{1}{c}{cadence} 
& \multicolumn{1}{c}{npts} 
& \multicolumn{1}{c}{time span} 
& \multicolumn{1}{c}{rms variance}\\ 
\multicolumn{1}{c} {} 
& \multicolumn{1}{c}{ } 
& \multicolumn{1}{c}{ } 
& \multicolumn{1}{c}{days} 
& \multicolumn{1}{c}{$\mu$mag}\\ 
\hline 
Q0  & LC & 471  & 9.71 & 1207 \\ 
Q1  & LC & 1621  & 33.45 & 1190 \\ 
Q3.3 & SC & 33\,679 & 25.07 & 1191 \\ 
\hline 
\end{tabular} 
\end{center} 
\end{table} 
 
Fig.\,\ref{fig:lc} shows the long cadence light curve for Q0 and Q1 together in 
the top panel and the short cadence light curve for Q3.3 in the middle panel. A 
few outlying points were edited from the raw data and a linear trend to correct 
for instrumental drift was removed from the Q1 data. The data are in {\it Kepler} 
magnitude, $Kp$, which is in broad-band white light. It is obvious from 
Fig.\,\ref{fig:lc} that KIC\,10195926 is an $\alpha^2$\,CVn star; i.e., it has 
surface spots that produce a variation in brightness with rotation. This is a 
common feature of the magnetic Ap stars and implies that the star has a strong, 
global magnetic field, even though that field is yet to be detected, and we were 
only able to place an upper limit on it of 5\,kG in 
Section\,\ref{sec:spectroscopy}. In some roAp stars, e.g. HR\,3831 
\citep{kochukhovetal04} and HD\,99563 \citep{freyhammeretal09},  the rare earth 
element surface spots are concentrated near to the pulsation or magnetic poles. 
These rare earth element spots often correlate with rotational light variations, 
particularly in the blue. With the advent of high precision space-based 
observations we can now see rotational light variations from spots in Ap stars 
that would not have been detected with ground-based observations, or at least 
would have given a poorly defined light curve. Such is the case for KIC\,10195926 
where the peak-to-peak range in $Kp$ is only slightly greater than 3\,mmag. We 
point out, however, that the rotational light variations of Ap stars are strongly 
wavelength dependent, as the result of flux redistribution caused by abundance 
spots, so that KIC\,10195926 may well show higher amplitude rotational light 
variations observed in filtered light such as Johnson $U$ or $B$. 
 
To determine the rotational period of KIC\,10195926 we co-added the 1-min 
integrations of Q3 to 30-min integrations and analysed all of the data listed in 
Table\,\ref{tab:journal}. There is a 158-d gap between the end of the Q1 data and 
the beginning of the Q3.3 data, giving rise to possible alias ambiguities in an 
analysis of the full data set. As can be seen from Fig.\,\ref{fig:lc} the 
rotational light curve is highly nonsinusoidal, so Fourier analysis is not the 
best choice for period determination. We chose to fit a harmonic series by linear 
least-squares to the data to search for the greatest reduction in the variance in 
the data. 
 
\begin{figure} 
\centering 
\includegraphics[width=8cm]{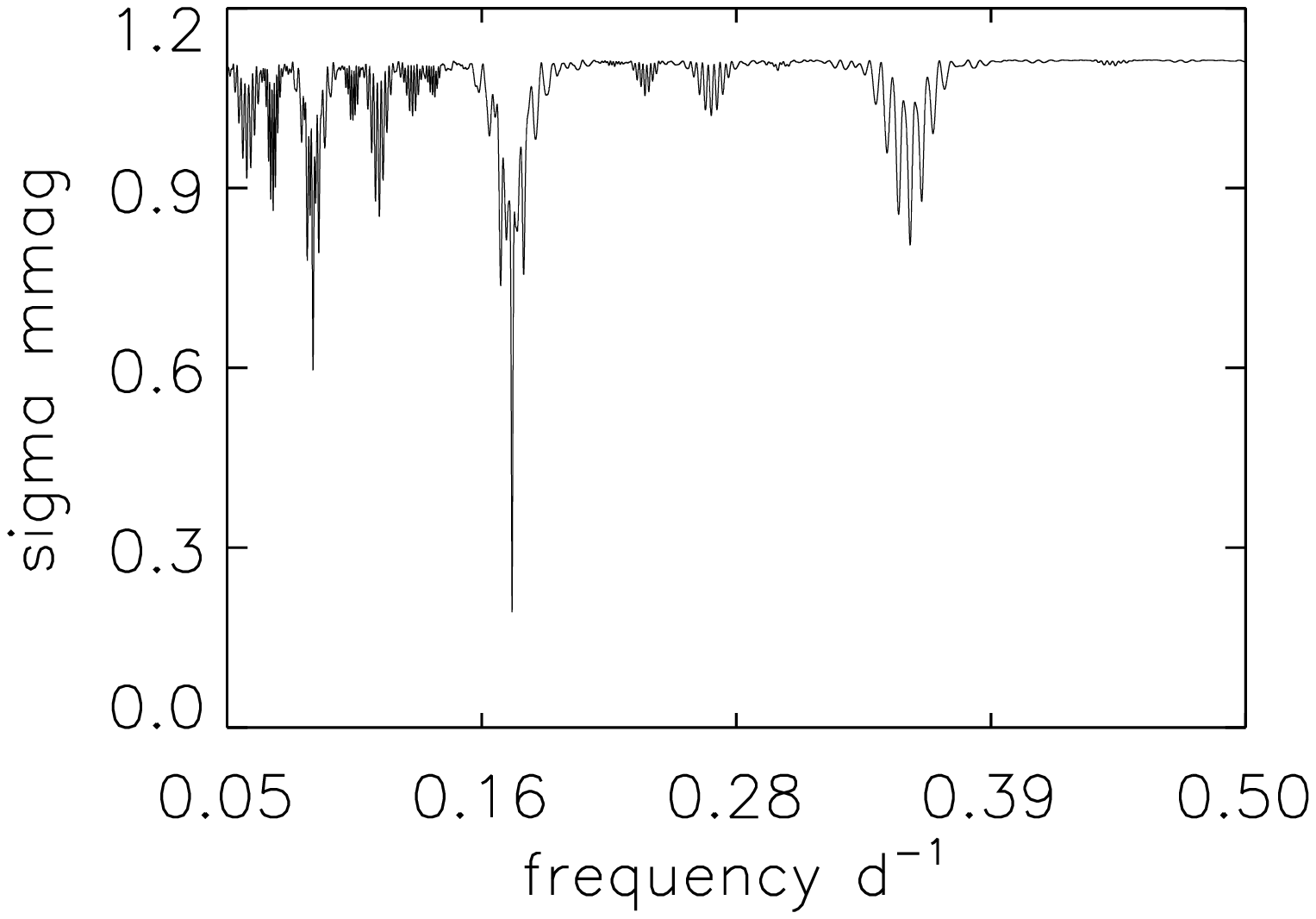} 
\includegraphics[width=8cm]{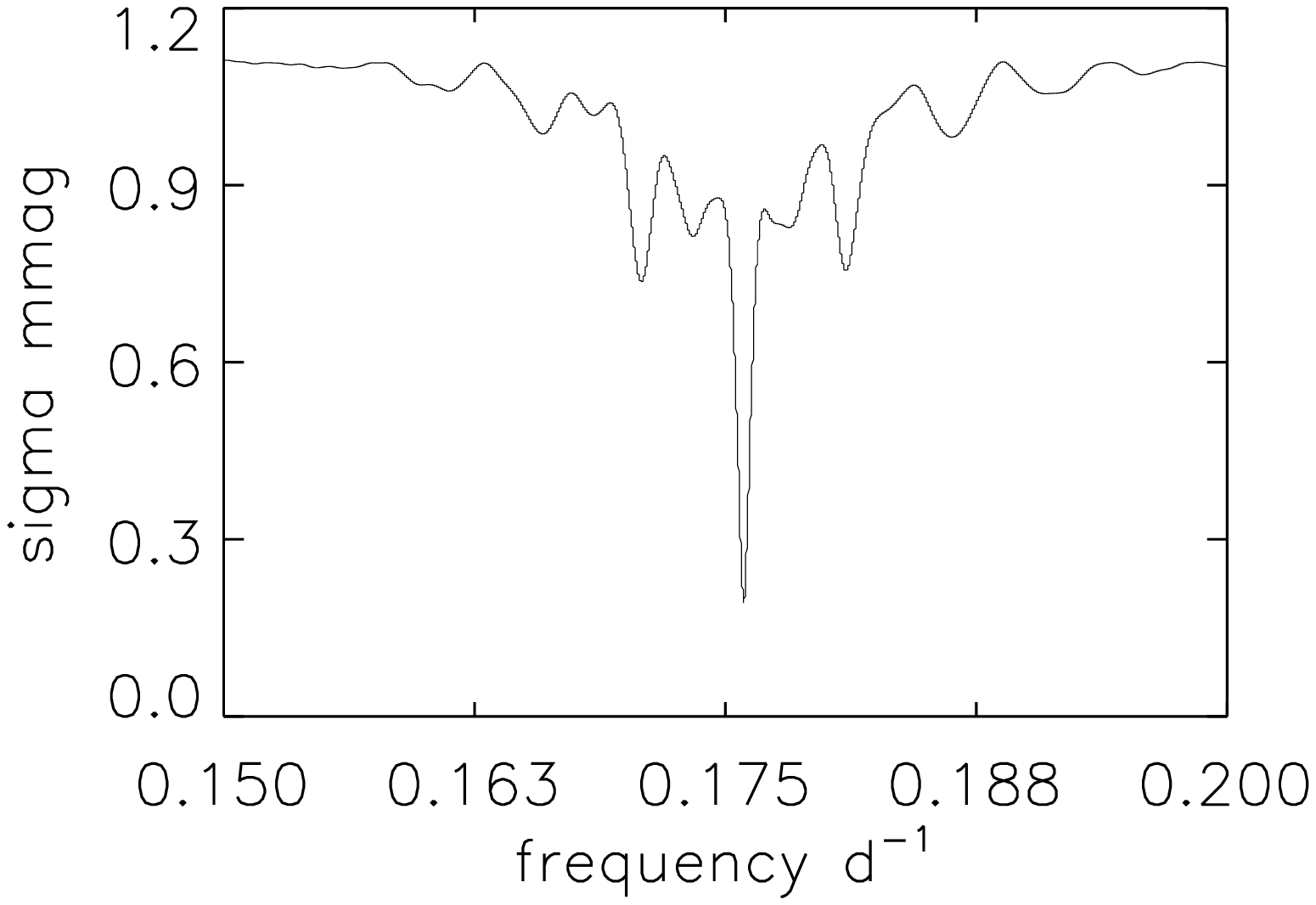} 
\caption{Top: This figure plots the standard deviation of one observation with 
respect to a fit of a fifth-order harmonic series at each frequency to the data. 
The procedure breaks down at low frequency, and is meaningless at zero frequency, 
hence the choice of frequency range searched. Bottom: This is a higher resolution 
view of the best-fitting frequency that shows there is no alias ambiguity, despite 
the 158-d gap in the data between Q1 and Q3.3. } 
\label{fig:lsps} 
\end{figure} 
 
\begin{figure} 
\centering 
\includegraphics[width=8cm]{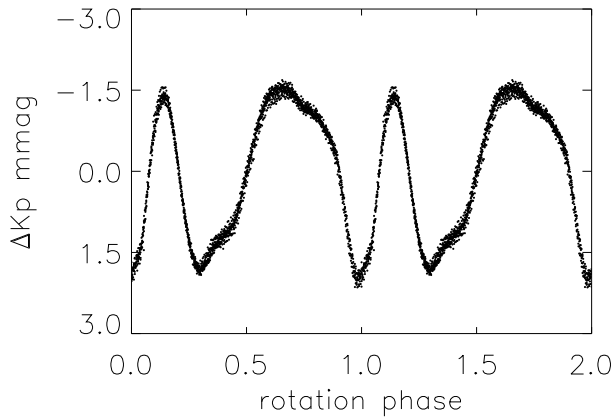} 
\caption{The rotational phase curve for the Q0 and Q1 data for KIC\,10195926. We 
have chosen the zero point of the time scale to be $t_0 = {\rm 
BJD}\,2455168.91183$, which is the time of pulsation maximum for the principal 
mode in this star. The diagram shows that pulsation maximum coincides with an 
extremum of the rotational light curve.} 
\label{fig:phase-q01} 
\end{figure} 
 
Fig.\,\ref{fig:lsps} shows the results of a fifth-order harmonic series fit to the 
data. There is a clear best frequency at $\nu_{\rm rot} = 0.17591$\,d$^{-1}$ with 
no alias ambiguity. While significant harmonics are present up to $15 - 20$ in the 
individual quarters of data, no discernible improvement in Fig.\,\ref{fig:lsps} 
can be seen with a higher-order fit. 
 
We then fitted 20 harmonics of $\nu_{\rm rot} = 0.17591$\,d$^{-1}$ to the full 
data set by least-squares and then nonlinear least-squares to optimise the 
frequency and make an error estimate. This gave our best determination of the 
rotation frequency of $\nu_{\rm rot} = 0.175914 \pm 0.000004$\,d$^{-1}$, which 
corresponds to a rotation period of $P_{\rm rot} = 5.68459 \pm 0.00013$\,d. This 
error estimate is internal. Some systematic effects are still present in the data, 
such as small instrumental drifts and slight changes in amplitude between 
quarters. These will be improved in future data reductions using more detailed 
pixel information. Since more data for KIC\,10195926 are being obtained, the 
rotation period will be determined to higher accuracy in future studies. The 
precision given here is excellent for comparison with the pulsation phases within 
the oblique pulsator model. 
 
Fig.\,\ref{fig:phase-q01} shows the long cadence light curve for Q0 and Q1 phased 
with the rotational period. The double-wave character of the rotational light 
curve indicates two principal spots. The zero point of the time scale has been 
selected to coincide with the time of pulsation maximum, as determined in 
Section\,\ref{sec:pulsations} below. It can be seen that minimum rotational 
brightness coincides with pulsation maximum, as is expected in the oblique 
pulsator model \citep{kurtz82}, in the case when the spots are associated with the 
magnetic/pulsation axis of the star. This coincidence can also be seen by 
comparing the middle and bottom panels of Fig.\,\ref{fig:lc}. 
 
\begin{figure} 
\centering 
\includegraphics[width=8cm]{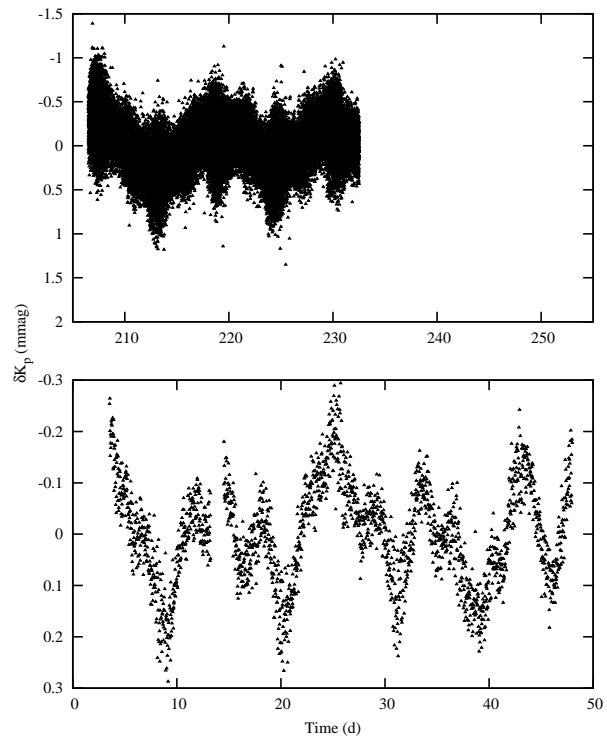} 
\caption{Light variation after removal of rotational modulation. The time scale is 
the same in both plots with time zero chosen to be BJD~2454950.000. The top panel 
shows the short-cadence data. Note the higher scatter at light minima and light 
maxima due to the roAp pulsation. The bottom panel shows the long-cadence data 
after correction of long-term drift. In both cases a clear variation with twice 
the rotational period can be seen. } 
\label{fig:lcurvepw} 
\end{figure} 
 
The asymmetry of the light curve requires differences in spots on opposite 
hemispheres. It does not appear that the spots seen at rotational phase 0.5 are 
coincident with the second pulsation pole. The spots of Ap stars are often much 
more pronounced than this in filtered light (say, $UBVRI$), but are generally not 
studied previously in white light. An exception is $\alpha$\,Cir, which was 
studied in white light with the star-tracker on the WIRE mission 
\citep{brunttetal09}. While the rotational light curve of $\alpha$\,Cir is more 
symmetrical, it has a similar low amplitude of only a few mmag. 
 
Interestingly, \citet{brunttetal09} showed that the pulsation pole for 
$\alpha$\,Cir is not coincident with the spots. As that is the case here also for 
the second spot at rotation phase 0.5 of KIC\,10195926, we speculate that as more 
precise rotational light curves are obtained for Ap stars, we will find more low 
amplitude rotational variation where the spots are not coincident with the 
magnetic poles. We also speculate that this will show a correlation with magnetic 
fields that are less strongly dipolar than in Ap stars with higher amplitude 
rotational light variations. 
 
\begin{figure} 
\centering 
\includegraphics[width=8cm]{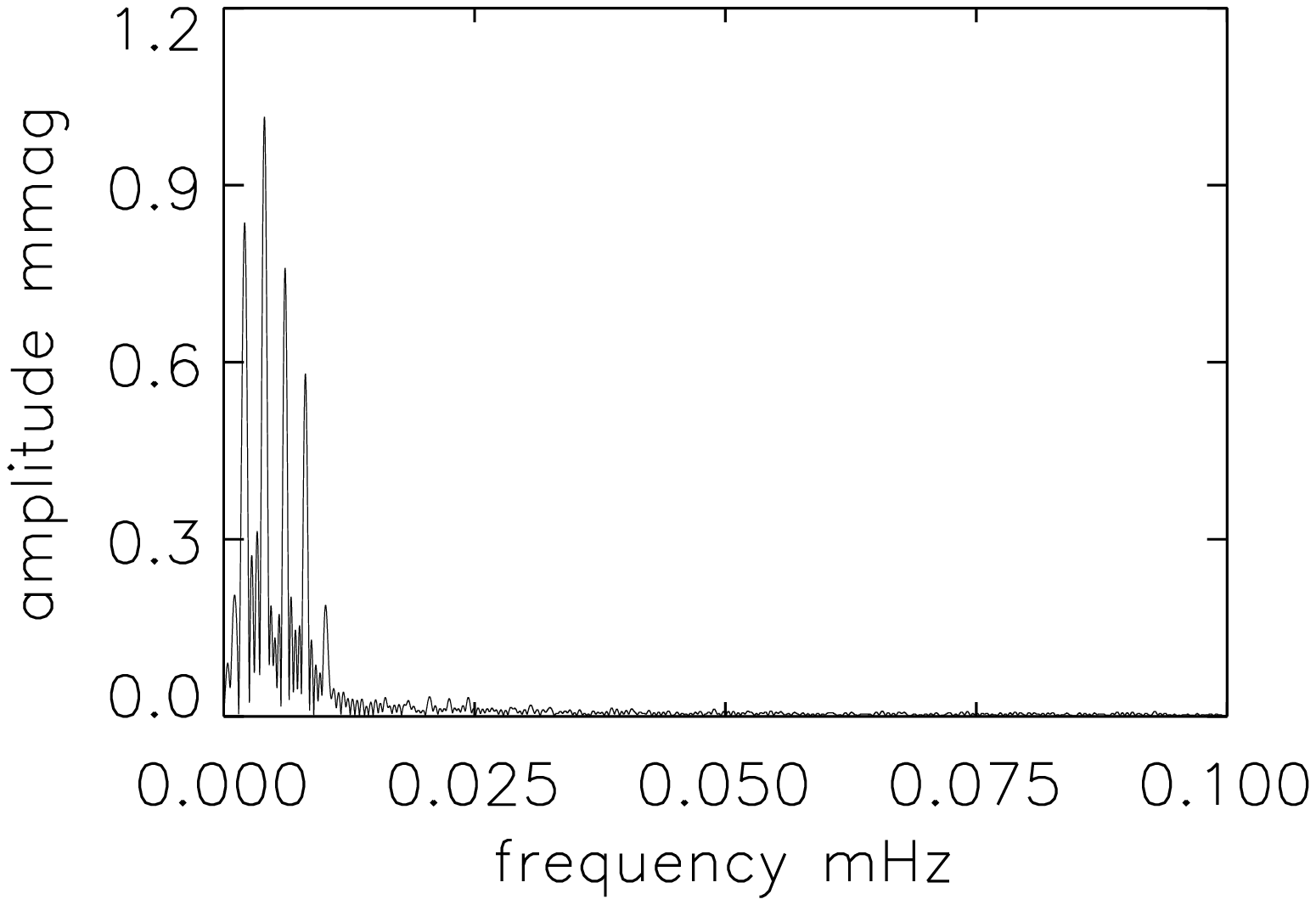} 
\includegraphics[width=8cm]{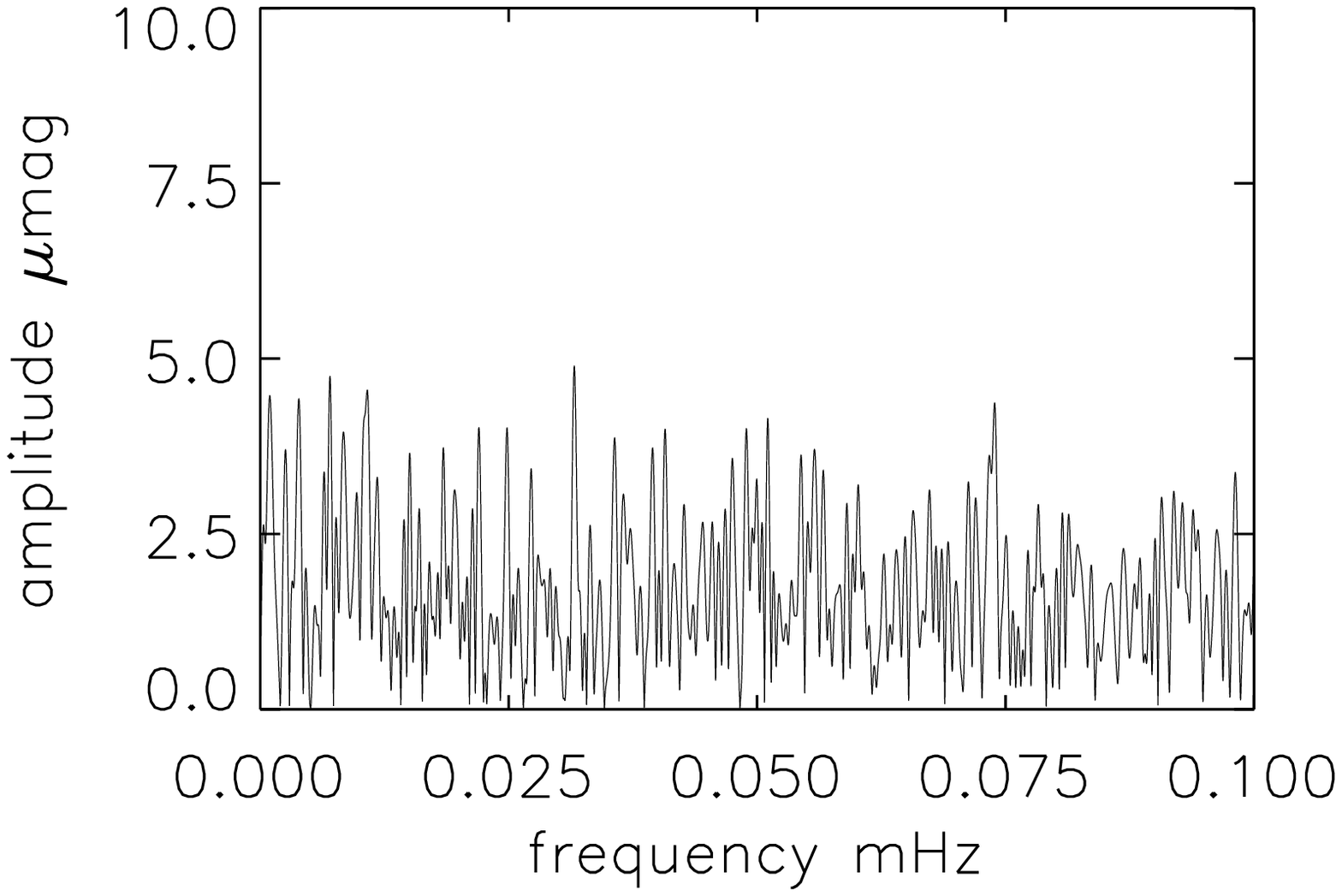} 
\caption{The upper panel is an amplitude spectrum of the Q3.3 short cadence data 
showing the low frequency spot variations. While it cannot be seen easily at this 
scale, the highest noise peaks have amplitudes less than 5\,$\mu$mag. A high pass 
filter was thus run to remove all variations at frequencies less than 0.05\,mHz 
with amplitudes greater than 5\,$\mu$mag. The bottom panel shows the amplitude 
spectrum in the same range after the high pass filtering. Note the change from 
mmag to $\mu$mag between these panels. } 
\label{fig:ft-lowfs} 
\end{figure} 
 
\subsection{The rotational subharmonic}
\label{sec:subharmonic}
 
When a $20^{\rm th}$-order harmonic series is fitted to the rotational light 
curves for the individual quarters, both Q0 and Q3.3 show a highly significant 
peak at $\frac{1}{2}\nu_{\rm rot}$, the subharmonic. That this appears in two 
independent data sets shows that it is not an artefact. The residual light 
variation after removal of the $20^{\rm th}$-order harmonic fit to the rotational 
light curve is shown in Fig.\,\ref{fig:lcurvepw} where it is evident that there is 
periodic variability on a time scale of $2 P_{\rm rot} = 11.3692$\,d. It is this 
that gives rise to the subharmonic and its harmonic in the frequency spectrum.

Ellipsoidal variables often have frequency spectra where the highest peak occurs 
at $2\nu_{\rm rot}$ as a consequence of a double-wave light curve with different 
maxima and minima. If the highest peak in the frequency spectrum were mistaken for 
$\nu_{\rm rot}$, then there would appear to be a subharmonic present. This is most 
unlikely to be the case for KIC\,10195926. The light curve seen in 
Fig.\,\ref{fig:lc} bears little resemblance to an ellipsoidal variable, and the 
incidence of short period binaries in magnetic Ap stars is very low. We are 
gathering high resolution spectra of this star to study its rotational variations. 
Those will be used to test for binary radial velocity variations. It is clear that 
the dominant rotational light variation is of the $\alpha^2$\,CVn type, i.e. that 
the star is a spotted magnetic variable. If the variation with $2P_{\rm rot}$ were 
to be interpreted as orbital, then the rotation period would have to be half that 
of the orbital period, which is again unlikely. 
 
Subharmonics are seen in stellar pulsation, most strikingly in the {\it Kepler} 
data for RR\,Lyr stars, including RR\,Lyrae itself (\citealt{szaboetal10}; 
\citealt{kolenbergetal10}). This is interpreted in terms of nonlinear effects in 
the pulsation, but this explanation does not apply here because we are dealing 
with rotation, not pulsation. 

We interpret the rotational light variations in Fig.\,\ref{fig:lc} to be caused by 
surface spots and we conclude that the rotation period of the star is $P_{\rm rot} 
= 5.68459$\,d. It is unlikely for an $\alpha^2$\,CVn spotted rotator to have two 
sets of spots that are as asymmetric as those of KIC\,10195926, but nearly 
identical on two opposite hemispheres, as would be required if the rotation period 
were twice the value given above. There is no known case of this amongst the 
$\alpha^2$\,CVn variables. 

We fitted by least-squares a harmonic series for $\frac{1}{2}\nu_{\rm rot}$. Other 
than $\nu_{\rm rot}$ and $\frac{3}{2}\nu_{\rm rot}$, only the harmonics of 
$\nu_{\rm rot}$ were significant. Thus if the subharmonic were the rotation 
frequency, we would have a harmonic series describing the data that included only 
every other harmonic. This, too, is unprecedented and unlikely.

\begin{figure} 
\centering 
\includegraphics[width=8cm]{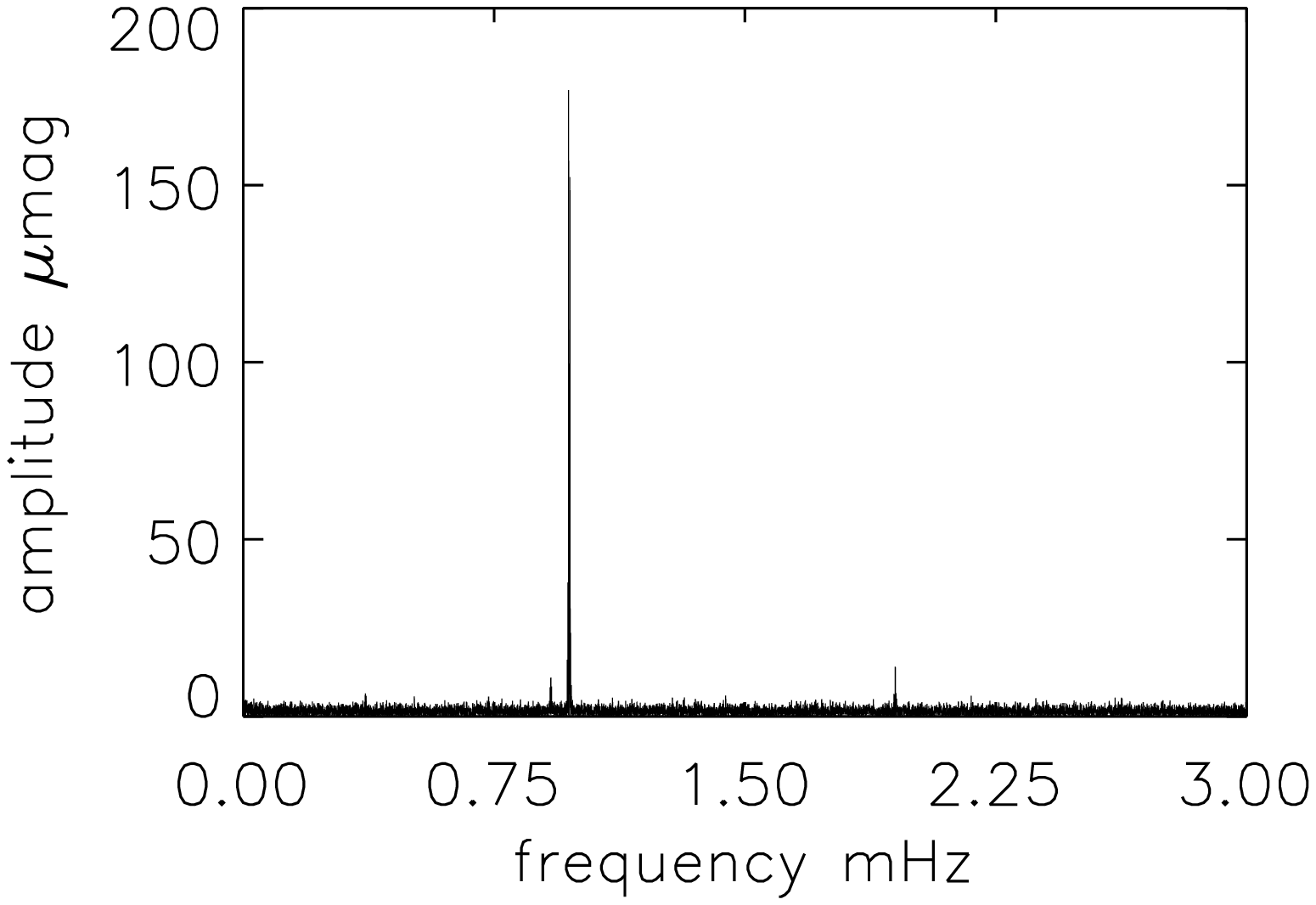} 
\includegraphics[width=8cm]{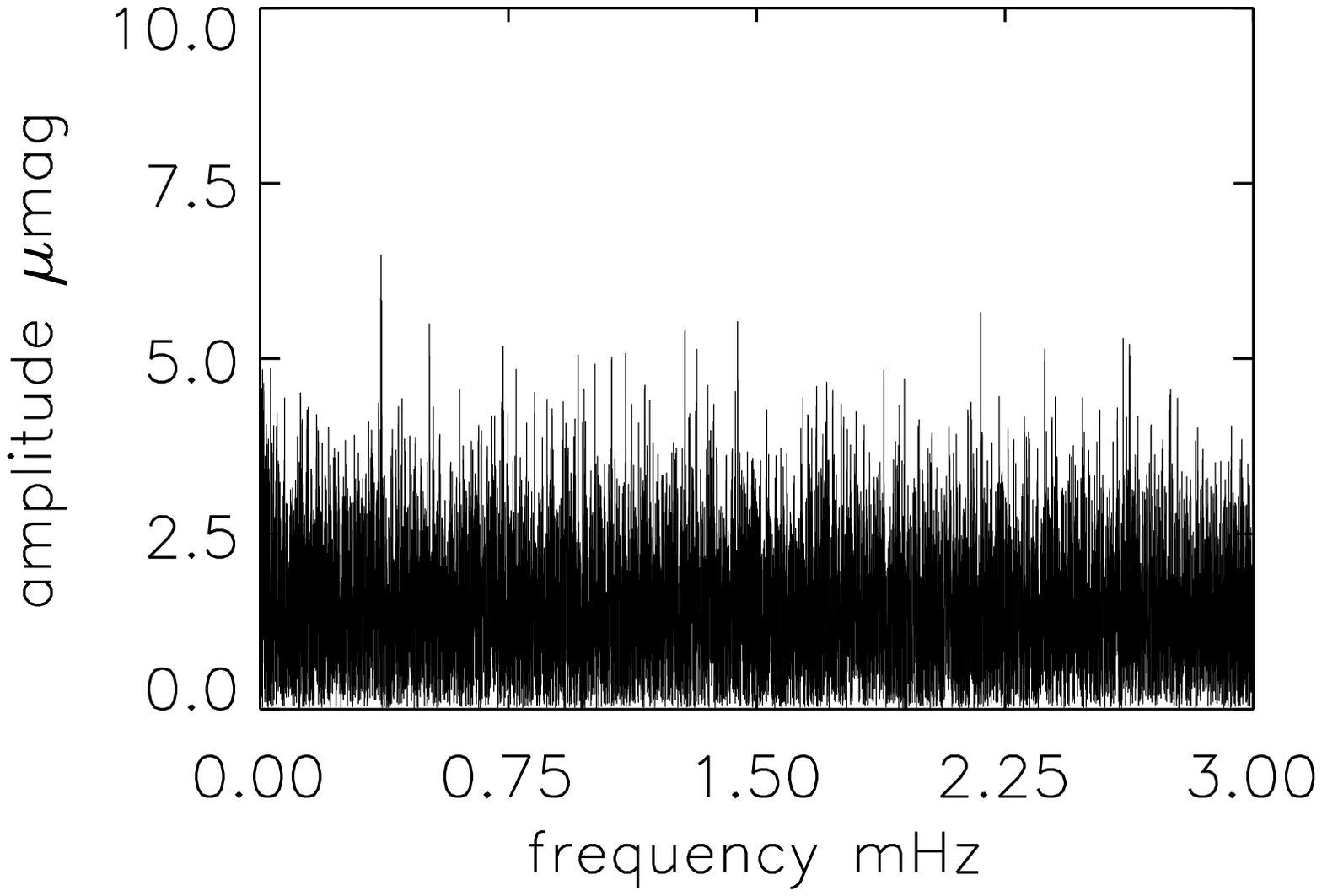} 
\caption{Top: An amplitude spectrum over the full range of pulsation frequencies. 
The principal frequency multiplet around $\nu_1$ and its harmonic can be seen, as 
well as the frequency of a second mode close to the principal at $\nu_2$. Note 
that the ordinate scale is in $\mu$mag. Bottom: An amplitude spectrum over the 
full range of pulsation frequencies for the residuals to the fit in 
Table\,\ref{tab:basicfit}. One significant frequency is present at $\nu_3 = 
0.365$\,mHz (31.54\,d$^{-1}$) with an amplitude of $6.5 \pm 1.2$\,$\mu$mag. This 
is formally significant at the 5.4$\sigma$ level, but it is a known {\it Kepler} 
artefact.} 
\label{fig:ft-full} 
\end{figure} 
 
\begin{figure} 
\centering 
\includegraphics[width=8cm]{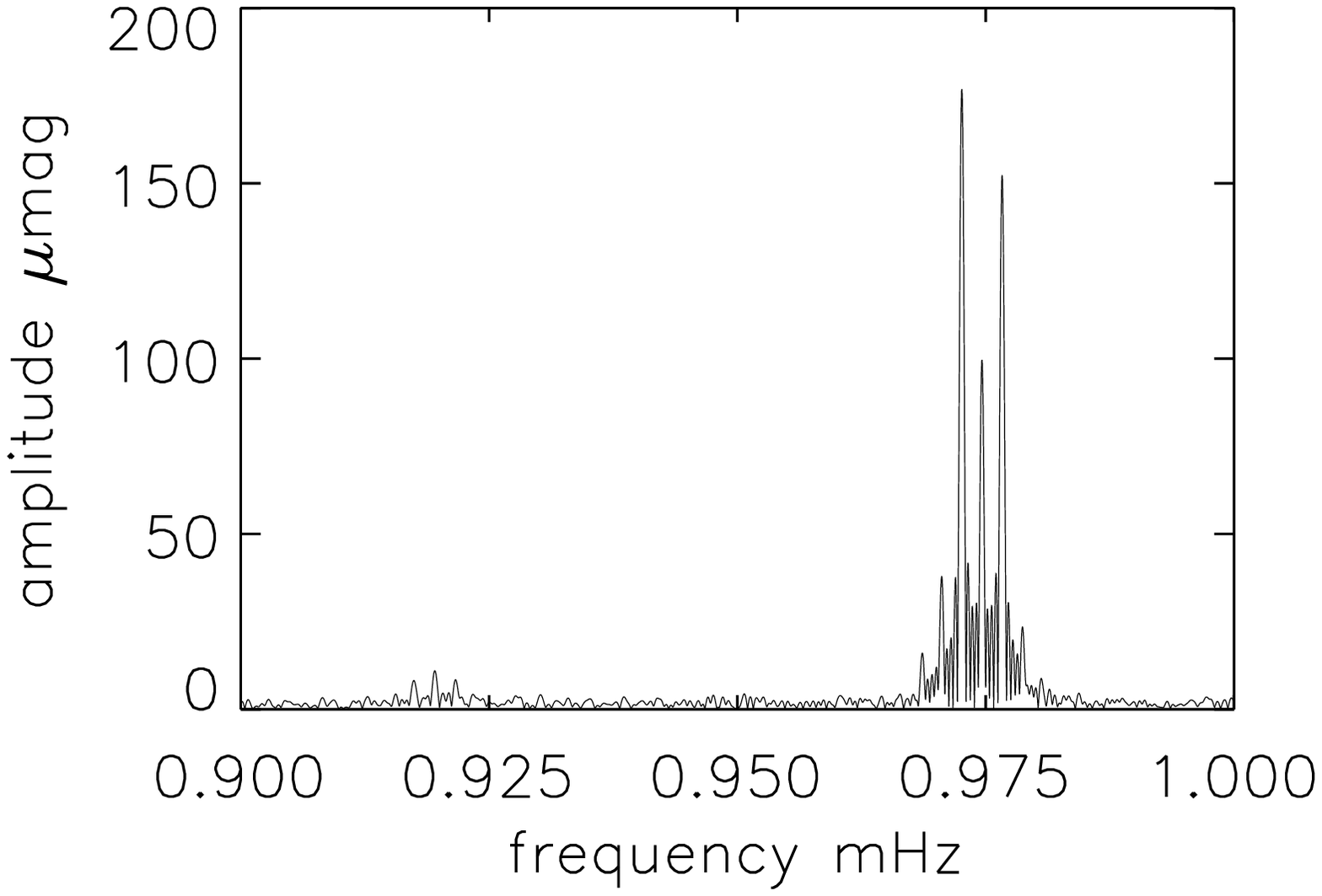} 
\includegraphics[width=8cm]{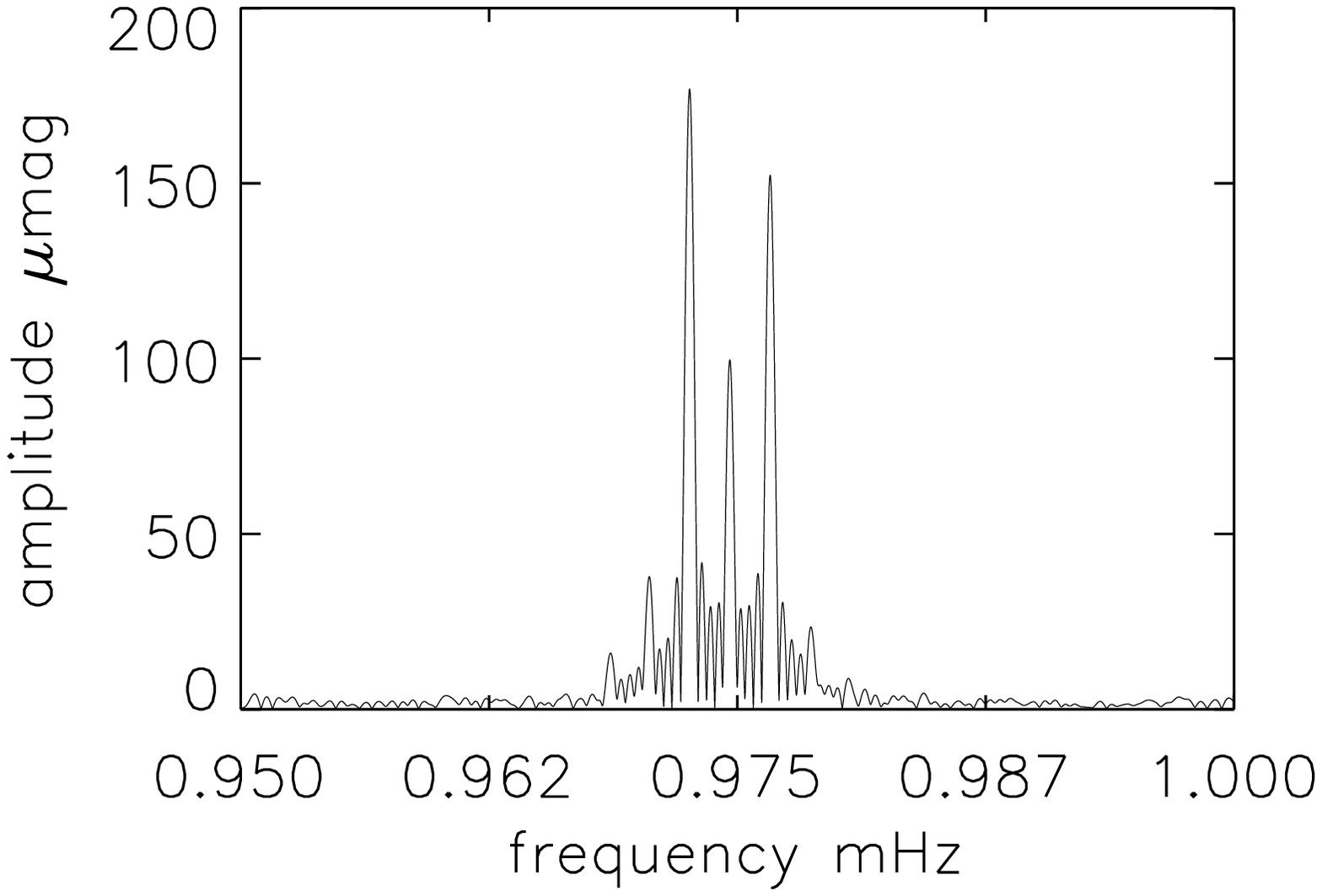} 
\includegraphics[width=8cm]{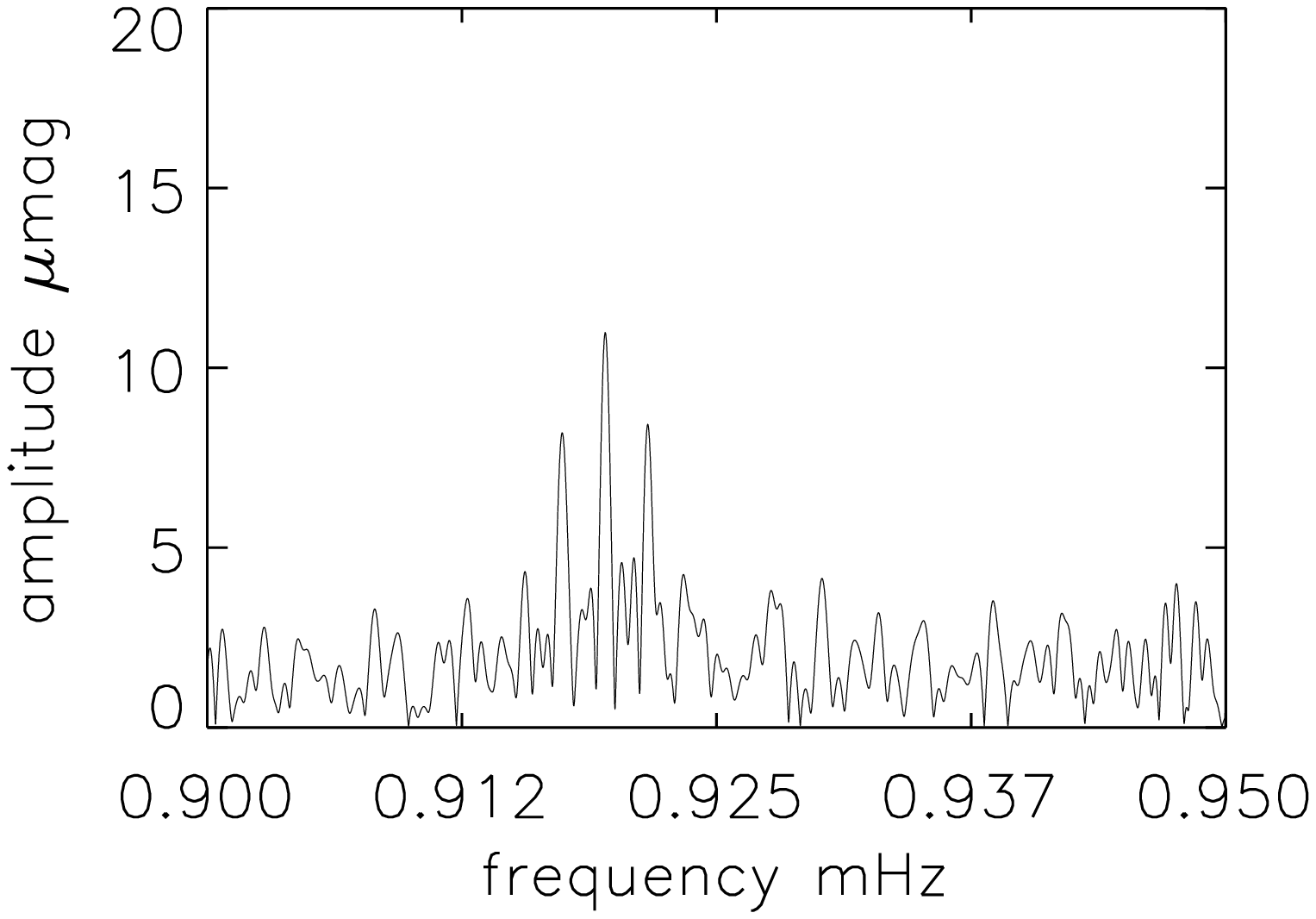} 
\caption{Amplitude spectra in the frequency range of $\nu_1$ and $\nu_2$. The top 
panel shows the multiplets of $\nu_1$ and $\nu_2$. The middle and bottom panels 
show them individually at higher resolution. Note the difference in the multiplet 
amplitude patterns, as this is significant for mode identification.} 
\label{fig:ampspec} 
\end{figure} 
 
\section{The pulsation frequencies} 
\label{sec:pulsations} 
 
Fig.\,\ref{fig:ft-lowfs} shows an amplitude spectrum of the short cadence data 
seen in the middle panel of Fig.\,\ref{fig:lc}. To study the pulsation frequencies 
we ran a high pass filter to remove completely the rotational light variations, 
and any instrumental drift. This was done by sequential automatic prewhitening of 
sinusoids until the noise level at low frequencies matches that at higher 
frequencies. The highest amplitude noise peaks are less than 5\,$\mu$mag, so all 
peaks with amplitudes greater than this at frequencies less than 0.05\,mHz were 
removed from the data. The purpose of this high pass filter is to ensure white 
noise so that the error estimates for the high frequency peaks are realistic. 
There is no significant crosstalk between the low frequency rotational variations 
and the high frequency pulsational variations. 
 
The bottom panel of Fig.\,\ref{fig:lc} shows these high-pass filtered data, where 
it can be seen that the pulsation amplitude is modulated twice per rotation 
period. This is because the principal pulsation mode is a distorted dipole mode 
and the rotational inclination, $i$, and magnetic obliquity, $\beta$, are such 
that $i + \beta > 90^\circ$, so that both poles are seen. That the maximum 
amplitude is different when alternating poles are presented suggests that neither 
$i$ nor $\beta$ is close to $90^\circ$. 
 
Fig.\,\ref{fig:ft-full} shows an amplitude spectrum over the full frequency range 
of variations seen in the star. The principal pulsation mode is seen at $\nu_1 = 
972.58$\,$\mu$Hz, and its harmonic can be seen at twice that frequency. Both of 
these are multiplets split by the rotation frequency. There is another pulsation 
mode close to $\nu_1$ at $\nu_2 = 919.55$\,$\mu$Hz. These are the second lowest 
frequencies (i.e., the second longest pulsation periods) detected in roAp stars, 
after HD\,116114 \citep{elkinetal05}, as a consequence of a relatively advanced 
evolutionary state of KIC\,10195926. We return to this point in 
Sections\,\ref{sec:secondary} and \ref{sec:HRD} (see Fig.\,\ref{fig:HRDiagram}). 
We also note that the bright roAp star $\beta$\,CrB \citep{kurtzetal07} has a 
similar low pulsation frequency and also shows overabundances of Nd\,\textsc{iii} 
and Pr\,\textsc{iii} of only 1\,dex as opposed to the typical 3\,dex seen in many 
roAp stars -- one of their characteristic signatures \citep{ryabetal04}. We return 
to this point in Section\,\ref{sec:discussion}. 
 
Fig.\,\ref{fig:ampspec} shows a higher resolution look at $\nu_1$ and $\nu_2$. The 
middle panel shows the multiplet of $\nu_1$. It is a frequency septuplet split by 
exactly the rotation frequency. This pattern is reminiscent of the well-studied 
roAp star HR\,3831 (see, e.g., \citealt{kurtzetal97}; \citealt{kurtzetal90}). It 
is an indication of a perturbed dipole mode for which $i + \beta > 90^\circ$, 
consistent with the rotational variation seen in Fig.\,\ref{fig:lc} where there is 
a double wave. The bottom panel of Fig.\,\ref{fig:ampspec} shows the frequency 
triplet of $\nu_2$. This is different to the pattern of $\nu_1$, as here the 
central peak has the highest amplitude. Thus $\nu_1$ and $\nu_2$ are either 
associated with modes of different degree, $l$, or have different pulsation axes. 
We return to this point in Section\,\ref{sec:opm}. 
 
\begin{figure} 
\centering 
\includegraphics[width=8cm]{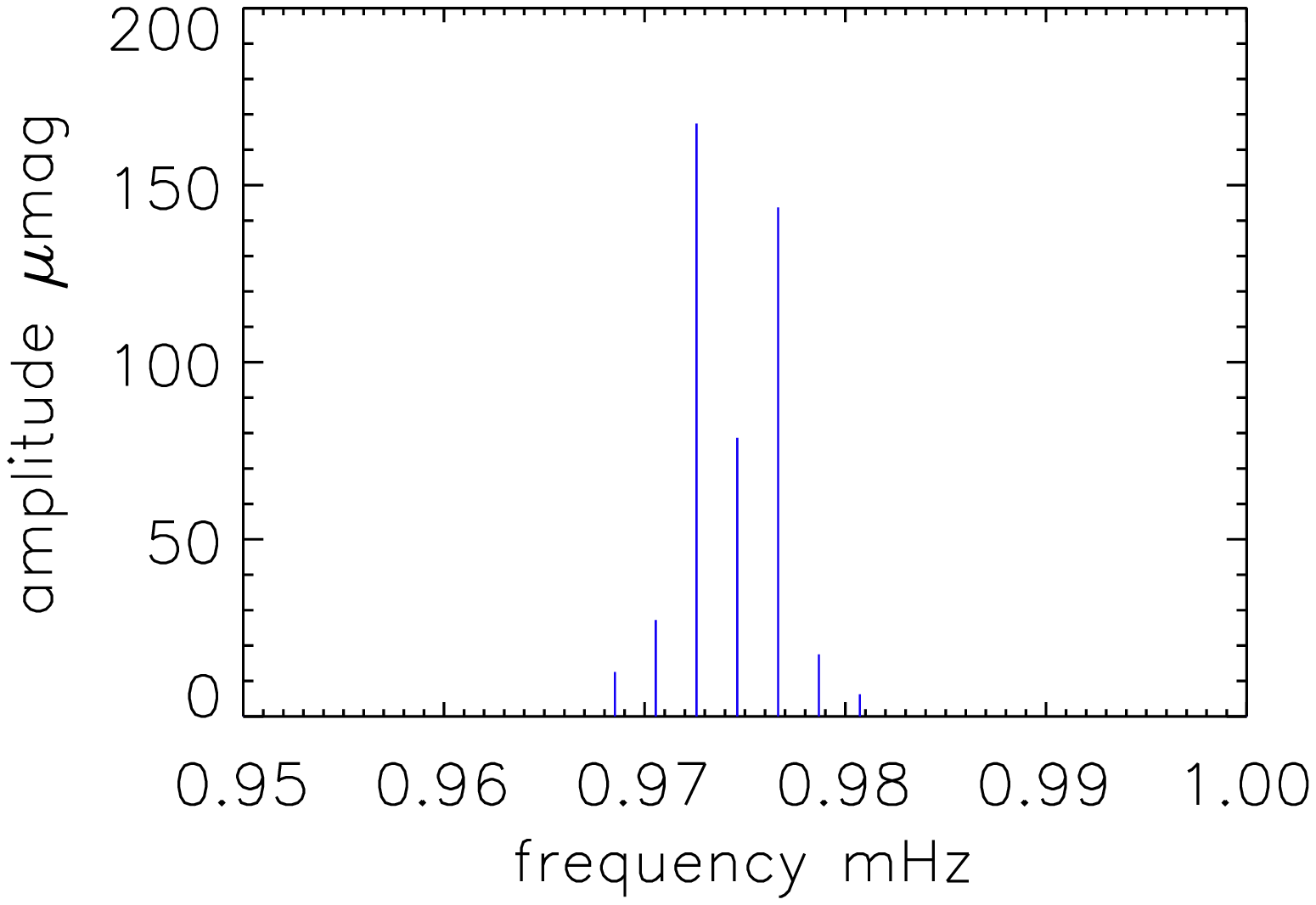} 
\includegraphics[width=8cm]{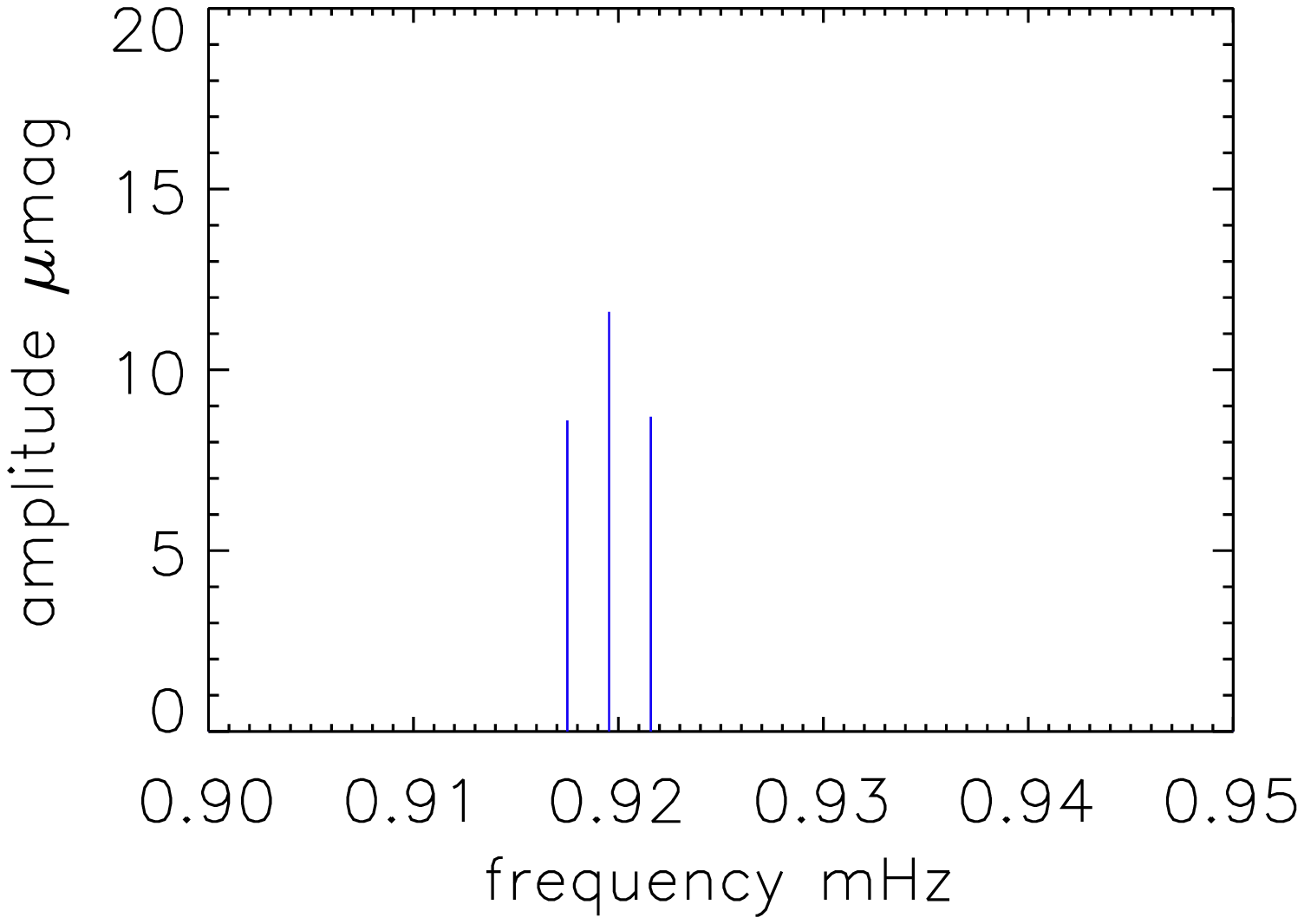} 
\caption{Schematic amplitude spectrum for the $\nu_1$ septuplet and $\nu_2$ 
triplet showing the component frequencies from the bottom two panels of 
Fig.\,\ref{fig:ampspec}. The splitting between the components of the multiplets is 
equal to the rotation frequency. The actual pulsation frequencies are those for 
the central peaks of the multiplets: $\nu_1 = 974.6183$\,$\mu$Hz and $\nu_2 = 
919.5464$\,$\mu$Hz.} 
\label{fig:schematic} 
\end{figure}

Table\,\ref{tab:basicfit} lists the values of the derived frequencies from a non-
linear least-squares fit to the data. We found the highest peak in the amplitude 
spectrum, then fitted it, along with all previously identified peaks, by linear 
least-squares to the data. We then prewhitened the data by that solution and 
searched the residuals for the next highest peak until the noise level was 
reached. After all of the frequencies shown in Table\,\ref{tab:basicfit} were 
found, they were then fitted by non-linear least-squares to the high-pass filtered 
data set. As the noise is white for these data, the formal errors are appropriate. 
After prewhitening by the solution in Table\,\ref{tab:basicfit}, there is one 
significant peak left at 31.5\,d$^{-1}$ -- a known artefact in the {\it Kepler}  
data -- as is shown in the bottom panel of Fig.\,\ref{fig:ft-full}. Otherwise 
there is nothing but white noise with highest peaks of 6\,$\mu$mag. 
Fig.\,\ref{fig:schematic} shows the frequency multiplets for $\nu_1$ and $\nu_2$ 
schematically. 
 
\begin{table*} 
\caption[]{A non-linear least-squares fit of the $\nu_1$ septuplet and its 
quintuplet harmonic, plus the $\nu_2$ triplet. The last column shows the frequency 
difference between the frequency on that line and the one on the previous line. 
This is the rotation frequency. The value of $\nu_{\rm rot} = 2.034 \pm 
0.005$\,$\mu$Hz translates to $P_{\rm rot} = 5.690 \pm 0.014$\,d, which is 
consistent with $P_{\rm rot} = 5.68459 \pm 0.00013$\,d derived in 
Section\,\ref{sec:lightcurves} from the rotational light variations. The zero 
point for the phases is BJD\,245\,5169.0. The residuals per observation for this 
fit equal 169.8\,$\mu$mag. } 
\label{tab:basicfit} 
\begin{center} 
\begin{tabular}{lrrrcc} 
\hline 
\multicolumn{1}{c}{ID} 
& \multicolumn{1}{c}{frequency} 
& \multicolumn{1}{c}{amplitude } 
& \multicolumn{1}{c}{phase} 
& \multicolumn{1}{c}{frequency difference}\\ 
\multicolumn{1}{c} {} 
& \multicolumn{1}{c}{$\mu$Hz} 
& \multicolumn{1}{c}{$\mu$mag} 
& \multicolumn{1}{c}{radians} 
& \multicolumn{1}{c}{$\mu$mag}\\ 
\hline 
$\nu_2 - \nu_{\rm rot}$  & $917.4581 \pm 0.0367$ & $8.7 \pm 1.3$ & $0.031  
\pm 0.143$ &  &  \\ 
$\nu_2$  & $919.5464 \pm 0.0278$ & $11.6 \pm 1.3$ & $2.232 \pm 0.108$ & $2.088  
\pm 0.046$ \\ 
$\nu_2 + \nu_{\rm rot}$  & $921.6244 \pm 0.0366$ & $8.8 \pm 1.3$ & $-1.797 \pm 
0.143$ & $2.078 \pm 0.046$ \\ 
$\nu_1 - 3\nu_{\rm rot}$  & $968.5151 \pm 0.0260$ & $12.5 \pm 1.3$ & $-1.803  
\pm 0.100$ &  &  \\ 
$\nu_1 - 2\nu_{\rm rot}$  & $970.5569 \pm 0.0121$ & $27.3 \pm 1.3$ & $-2.308  
\pm 0.046$ & $2.042 \pm 0.029$ \\ 
$\nu_1 - \nu_{\rm rot}$  & $972.5845 \pm 0.0020$ & $167.5 \pm 1.3$ & $-1.757  
\pm 0.008$ & $2.028 \pm 0.012$ \\ 
$\nu_1$  & $974.6183 \pm 0.0043$ & $78.6 \pm 1.3$ & $-1.438 \pm 0.016$ & 
$2.034 \pm 0.005$ \\ 
$\nu_1 + \nu_{\rm rot}$ & $976.6569 \pm 0.0023$ & $143.7 \pm 1.3$ & $-1.563  
\pm 0.009$ & $2.039 \pm 0.005$ \\ 
$\nu_1 +2\nu_{\rm rot}$ & $978.7346 \pm 0.0181$ & $17.9 \pm 1.3$ & $-0.390 \pm 
0.070$ & $2.078 \pm 0.018$ \\ 
$\nu_1 + 3\nu_{\rm rot}$ & $980.7187 \pm 0.0520$ & $6.2 \pm 1.3$ & $-0.968 \pm 
0.201$ & $1.984 \pm 0.055$ \\ 
$2\nu_1 - 2\nu_{\rm rot}$ & $1945.0322 \pm 0.0513$ & $6.3 \pm 1.3$ & $2.462  
\pm 0.199$ &  &  \\ 
$2\nu_1 - \nu_{\rm rot}$ & $1947.2577 \pm 0.0666$ & $4.9 \pm 1.3$ & $2.367 \pm 
0.255$ & $2.226 \pm 0.084$ \\ 
$2\nu_1$ & $1949.2662 \pm 0.0246$ & $13.4 \pm 1.3$ & $2.777 \pm 0.094$ & 
$2.009 \pm 0.071$ \\ 
$2\nu_1 + \nu_{\rm rot}$ & $1951.2577 \pm 0.0564$ & $5.8 \pm 1.3$ & $-2.740  
\pm 0.217$ & $1.991 \pm 0.062$ \\ 
$2\nu_1 +2\nu_{\rm rot}$ & $1953.2500 \pm 0.0753$ & $4.3 \pm 1.3$ & $2.942 \pm 
0.290$ & $1.992 \pm 0.094$ \\ 
\\ 
\hline 
$\nu_1 - \nu_2$ & $55.072 \pm 0.028$ & & & &\\ 
\hline 
\end{tabular} 
\end{center} 
\end{table*} 

\subsection{The principal pulsation mode and the oblique pulsator model}
\label{sec:principal}

The separation of all the multiplets is consistent with the rotation frequency 
determined from the rotational light variations, as is expected in the oblique 
pulsator model. We therefore adopt the rotational frequency determined in 
Section\,\ref{sec:lightcurves}, which is $\nu_{\rm rot} = 2.03604 \pm 
0.00005$\,$\mu$Hz. We tested the hypothesis that the rotation period may be 
$2P_{\rm rot}$, by fitting by least-squares a 13-multiplet split by the 
subharmonic frequency, $\frac{1}{2}\nu_{\rm rot}$, as the $\nu_1$ multiplet. None 
of the additional peaks between those seen in Fig.\,\ref{fig:schematic} has an 
amplitude more than 2.5\,$\mu$mag, hence all lie well below the highest noise 
peaks of 6\,$\mu$mag. There is thus nothing in the pulsation modulation with 
rotation to suggest that $2P_{\rm rot}$ might be the rotation period. 

The frequency septuplet of $\nu_1$ of KIC\,10195926 is very similar to that of 
HR\,3831 \citep{kurtzetal90, kurtzetal97}, which is interpreted as arising from a 
distorted dipole mode. \citet{kurtz92} showed that the mode in HR\,3831 is 
primarily an oblique dipole mode from a spherical harmonic decomposition of its 
frequency septuplet. \citet{kochukhov04} found that both dipole and octupole 
components are needed to describe the radial velocity variations in HR\,3831. 
These results do not imply that both dipole and octupole {\it modes} are excited. 
There is only one mode, but the magnetic distortion of that mode from a normal 
mode can be described with multiple spherical harmonic components. 
\citet{dziembowski96}, \citet{bigotetal00}, \citet{cunha00} and \citet{saio05} 
show theoretically how the effect of a dipole magnetic field leads to this 
distortion of the pulsation mode. Saio (2005 -- his Fig.\,12) specifically shows a 
comparison of the latitudinal and horizontal displacements from his model with the 
observational results of \citet{kochukhov04} for HR\,3831. 

\begin{figure} 
\centering
\includegraphics[width=8cm]{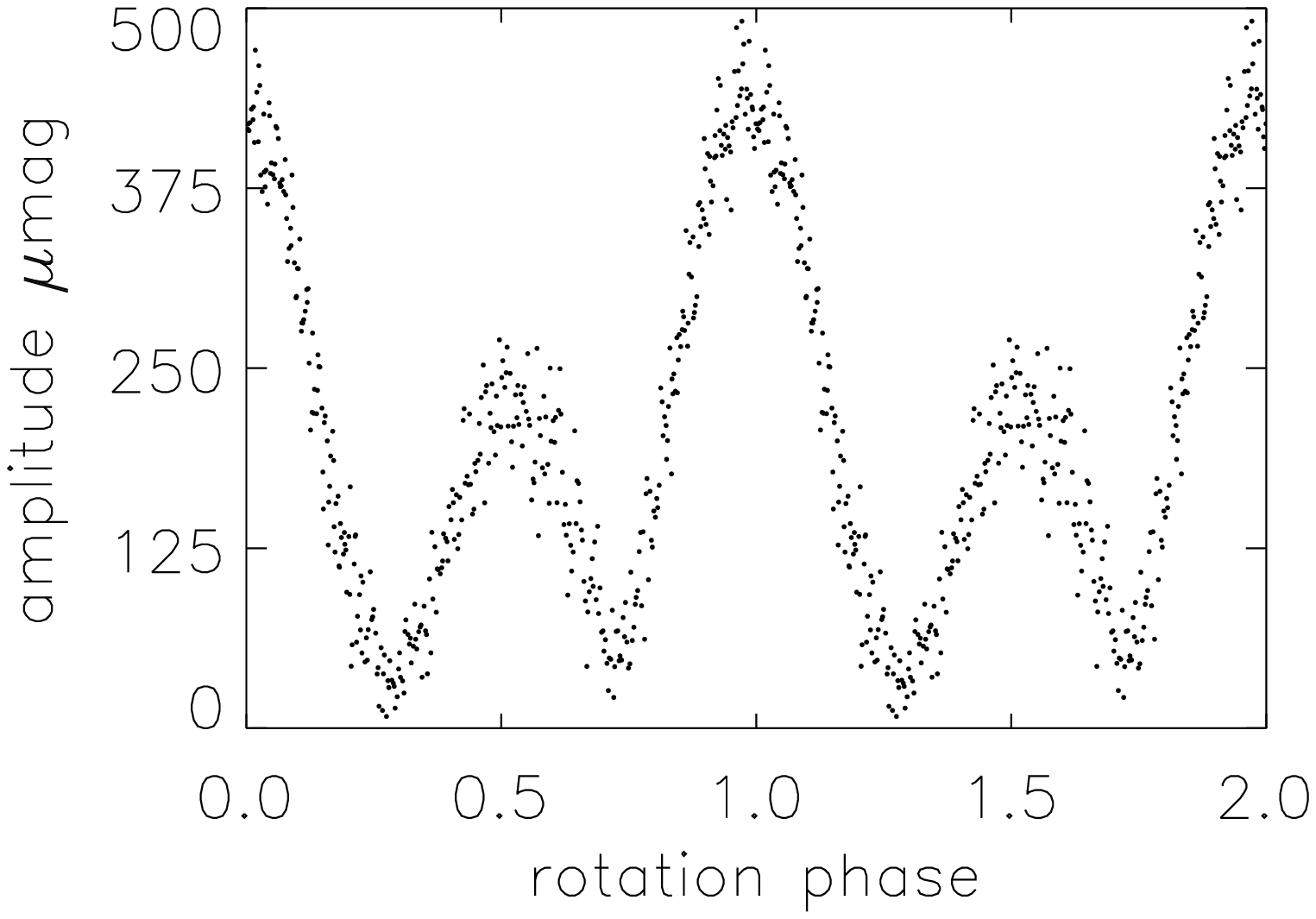} 
\includegraphics[width=8cm]{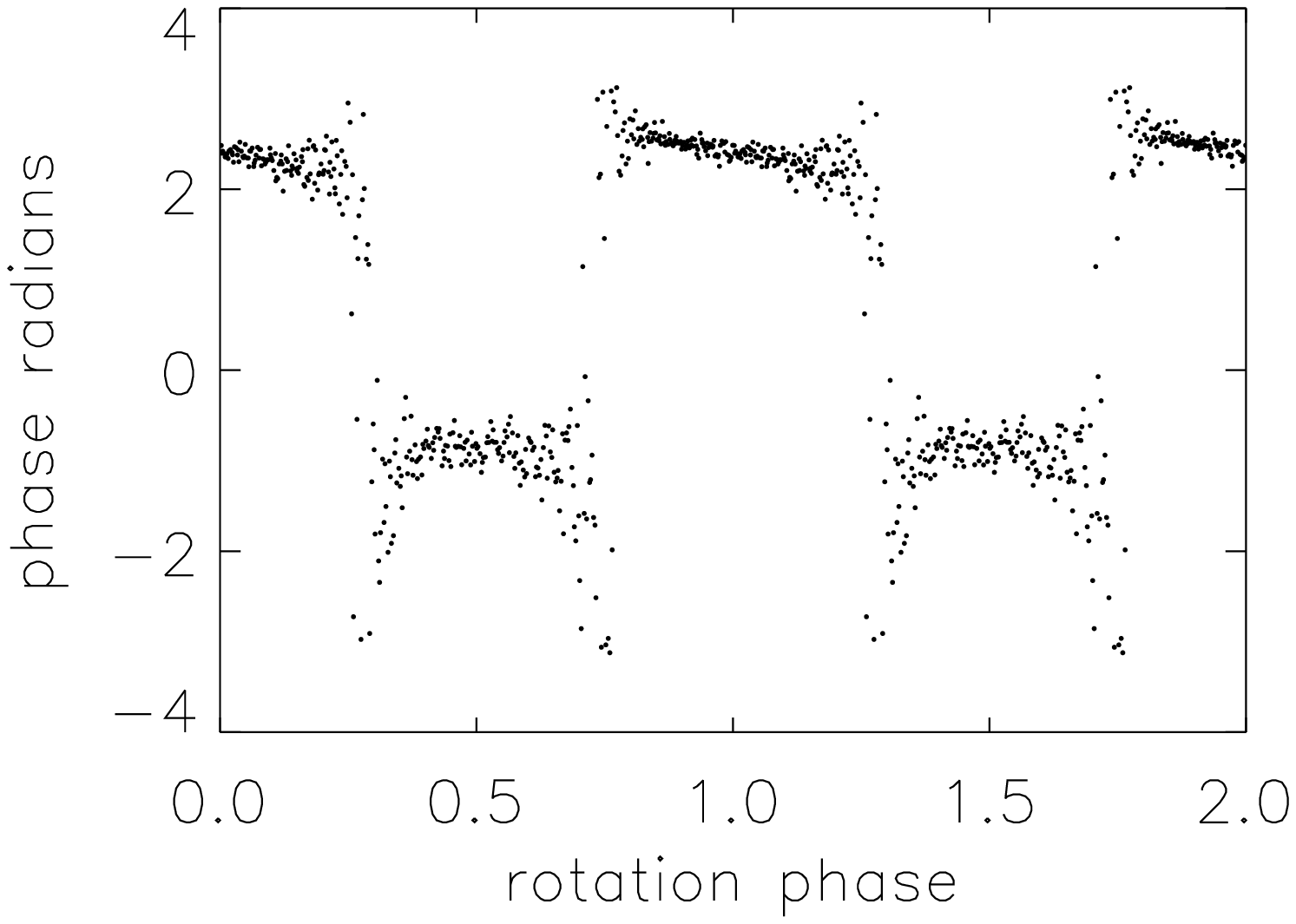} 
\caption{This shows the amplitude and phase modulation as a function of rotational 
phase for $\nu_1$. It has been created by fitting 5 cycles of data by linear least 
squares for $\nu_1$. Notice the $\pi$ radian reversal at quadrature; this is a 
typical sign of dipole pulsation in an oblique pulsator. The difference in 
amplitude of the two maxima is a function of the different aspects as the 
pulsation pole crosses the stellar meridian. Rotation phase zero is the same as in 
Fig.\,\ref{fig:phase-q01}, .i.e., with respect to $t_0 = {\rm 
BJD}\,2455168.91183$.} 
\label{fig:phamp1}
\end{figure}

However, as in the well known roAp stars HR\,3831, HR\,1217 and $\alpha$\,Cir, the 
frequencies of the modes in KIC\,10195926 appear dominantly to be triplets in 
photometry. Whether the latitudinal magnetic distortion is weak or strong, the 
mode geometry can be reasonably described by a dipole mode, since the higher order 
components (such as $l = 3$) have low visibility when averaged over the disk. For 
a dipole triplet, the phases of all three frequencies should be equal at the time 
of pulsation maximum for a simple oblique dipole mode, i.e. $\phi_{-1} = \phi_{+1} 
= \phi_{0}$ at that time. In Table\,\ref{tab:phasefit} we have chosen $t_0 = {\rm 
BJD}\,245\,5168.91183$ to force $\phi_{-1} = \phi_{+1}$. We then note that 
$\phi_{0}$ differs from the value of $\phi_{-1}$ and $\phi_{+1}$ by only $0.22 \pm 
0.02$\,rad. While this is significant (and is a result of the magnetic distortion 
of the mode), it is close enough to support an approximation of the central 
triplet as arising from a simple oblique dipole mode. We also note that 
Fig.\,\ref{fig:phase-q01} shows that this time corresponds to the rotational light 
minimum, showing that the abundance spot that gives rise to this minimum is in the 
plane defined by the rotation and pulsation poles. This is confirmed in the phase 
curve of the principal pulsation mode $\nu_1$ shown in Fig.\,\ref{fig:phamp1}. We 
found the amplitude and phase of this mode by fitting $\nu_1$ to consecutive 5 
cycles of data. We have plotted that against rotation phase using the same zero 
point as in Fig.\,\ref{fig:phase-q01}. The amplitude modulation and $\pi$-radian 
phase reversal are characteristic of an oblique dipole pulsation mode (Kurtz 
1982). 

The magnetic field present in roAp stars partially raises the $(2l+1)$ degeneracy 
of non-magnetic modes whose properties (frequency, amplitude, phase) then depend 
on the absolute value of their azimuthal order $m$. The modes can be either 
aligned or perpendicular to the field axis. However, in roAp stars, rotation is 
enough to play an important role in the mode properties. Indeed, since the 
magnetic and rotation axes are in general not aligned ($\beta\neq 0$), the 
rotation acts as an asymmetric perturbation on the magnetic modes, which are 
thereby coupled with respect to the azimuthal order (rotation is not strong enough 
in roAp stars to couple modes of different degrees $l$). 

In this case of rotationally coupled magnetic modes the dipole mode displacement 
vector describes an ellipse during the pulsation period which is defined by its 
two axes: the X-axis in the plane defined by the rotation and magnetic axes and 
the other, the Y-axis, perpendicular to this plane. This is the generalized 
oblique pulsator model presented by \citet{bigot02}. The mode is then fully 
described by the inclination of the ellipse that we defined here by the 
inclination $\gamma$ of the X-axis with respect to the rotation axis and the angle 
$\psi$ whose tangent is the ratio between the lengths of the Y- and X- axes; see 
\citet{bigot02} and \citet{bigot10} for more details. For a pure $m=0$ dipole mode 
in the magnetic reference system we have $\gamma=\beta$ and $\psi=0$. For a pure 
$m=\pm 1$ mode in the same frame we have $\gamma=\pi/2+\beta$ and $\psi=\pm 
\pi/4$. The light curves (or the triplet in the frequency domain) are described by 
only three angles: $i,\gamma,\psi$ (\citealt{bigot10}). The last two angles are 
functions of the magnetic field, rotation, $\beta$ and frequency. 

The strength of the centrifugal force coupling of the magnetic modes depends on 
the frequency difference between the magnetic modes $m=0$ and $m=\pm 1$ and the 
centrifugal shift of frequency. In order to measure the effects, \citet{bigot02} 
defined the parameter $\left | \mu^{\rm mag} \right |$ to be the ratio between the 
magnetic frequency shift divided by the centrifugal shift: $\left | \mu^{\rm mag} 
\right | < 1$ corresponds to a dominating rotation regime, $\left | \mu^{\rm mag} 
\right | = 1$ corresponds to strong coupling, $\left | \mu^{\rm mag} \right | > 1$ 
corresponds to a magnetic regime where there is almost no coupling. The Coriolis 
force is far weaker than the centrifugal force in roAp stars and almost does not 
enter into the balance that defines the inclination of the mode. However, it plays 
a role in the asymmetry of the amplitudes $A_{+1}$ and $A_{-1}$.

We do not know the magnetic field ($B_p,\beta$) configuration for this star; thus 
far we have only been able to limit the polar field strength to be less than 
5\,kG. Because rotation is not strong enough to couple modes of different $l$, the 
presence of latitudinal distortion of the mode, seen as a septuplet for $\nu_1$, 
is only due to the magnetic field and can be used as a constraint on the field 
strength. This is what is done in the following Section\,\ref{sec:models}. The 
effect of rotation is taken into account in Section\,\ref{sec:opm}. 

\begin{table} 
\caption[]{A linear least-squares fit of the $\nu_1$ septuplet and its quintuplet 
harmonic, plus the $\nu_2$ triplet. The rotation frequency has been forced to be 
equal to 2.034\,$\mu$Hz for the splittings of the multiplets. This is to test the 
phase relationships. The time zero point is BJD\,245\,5168.91183; it has been 
chosen to force the phases $\phi(\nu_1 - \nu_{\rm rot}) = \phi(\nu_1 + \nu_{\rm 
rot})$, Notably, the phase $\phi(\nu_1$) is close to the same value. Exact 
equality and only a frequency triplet is what is expected for a pure oblique 
dipole with no magnetic perturbation. The residuals per observation for this fit 
equal 169.8\,$\mu$mag. } 
\label{tab:phasefit}
\begin{center} 
\begin{tabular}{lrrr} 
\hline 
\multicolumn{1}{c}{ID} 
& \multicolumn{1}{c}{frequency} 
& \multicolumn{1}{c}{amplitude } 
& \multicolumn{1}{c}{phase} \\
\multicolumn{1}{c} {} 
& \multicolumn{1}{c}{$\mu$Hz} 
& \multicolumn{1}{c}{$\mu$mag} 
& \multicolumn{1}{c}{radians} \\
\hline 
$\nu_2 - \nu_{\rm rot}$  &$917.5104$ & $8.6 \pm 1.3$ & $0.120 \pm 0.145$ \\
$\nu_2$          &$919.5464$ & $11.6 \pm 1.3$ & $2.200 \pm 0.108$ \\
$\nu_2 + \nu_{\rm rot}$  &$921.5824$ & $8.7 \pm 1.3$ & $-1.964 \pm 0.144$ \\
 $\nu_1 - 3\nu_{\rm rot}$ &$968.5102$ & $12.4 \pm 1.3$ & $2.110 \pm 0.100$ \\
 $\nu_1 - 2\nu_{\rm rot}$ &$970.5462$ & $27.2 \pm 1.3$ & $1.503 \pm 0.046$ \\
 $\nu_1 -\nu_{\rm rot}$  &$972.5822$ & $167.4 \pm 1.3$ & $1.955 \pm 0.007$ \\
 $\nu_1$          &$974.6183$ & $78.6 \pm 1.3$ & $2.176 \pm 0.016$ \\
 $\nu_1 + \nu_{\rm rot}$  &$976.6543$ & $143.8 \pm 1.3$ & $1.956 \pm 0.009$ \\
 $\nu_1 + 2\nu_{\rm rot}$ &$978.6904$ & $17.6 \pm 1.3$ & $3.041 \pm 0.071$ \\
 $\nu_1 + 3\nu_{\rm rot}$ &$980.7264$ & $6.2 \pm 1.3$ & $2.343 \pm 0.203$ \\
 $2\nu_1 - 2\nu_{\rm rot}$ &$1945.1645$ & $5.6 \pm 1.3$ & $-2.893 \pm 0.222$ \\
 $2\nu_1 - \nu_{\rm rot}$ &$1947.2005$ & $4.9 \pm 1.3$ & $-2.772 \pm 0.258$ \\
 $2\nu_1$         &$1949.2366$ & $13.1 \pm 1.3$ & $-2.551 \pm 0.095$ \\
 $2\nu_1 + \nu_{\rm rot}$ &$1951.2726$ & $5.7 \pm 1.3$ & $-1.891 \pm 0.221$ \\
 $2\nu_1 + 2\nu_{\rm rot}$ &$1953.3087$ & $4.4 \pm 1.3$ & $-2.573 \pm 0.286$ \\
\hline
\hline
\end{tabular} 
\end{center} 
\end{table} 

\begin{table} 
\caption[]{We list all possible values of $(i,\beta)$ that lead to an inclination 
of the dipole mode $\nu_1$ in agreement with the amplitude ratio defined by the 
data in Table\,\ref{tab:phasefit}, i.e. $(A_{+1}+ A_{-1})/A_0 = 3.96 \pm 0.07$. We 
assume equal magnetic and centrifugal strength $\left | \mu^{\rm mag} \right | = 
1$. Columns 1 and 2 list the values of $i$ and $\beta$, respectively. Column 3 
lists the value of the inclination $\gamma$. Column 4 lists the inclination of the 
dipole axis with respect to the magnetic field. Column 5 lists the corresponding 
polarization ($\psi=0$ is linear). Columns 6 and 7 list the angles made by the 
pulsation axis with the line of sight at the time of amplitude maxima. Column 8 
lists the quantity $\cal{D}$ defined in \citet{bigot02} to measure the alignment 
with the magnetic field axis ($\cal{D} = {\rm 0}$ means alignment). 
The rows in bold characters corresponding to $i=[62^{\circ},65^{\circ}]$ lead to 
an amplitude ratio $(A_{+1}-A_{-1})/(A_{+1}+A_{-1})$  equal to the observed one. 
All angles are given in degrees.}
\label{tab:angles}
\begin{center} 
\begin{tabular}{rrrrrrrr} 
\hline 
\multicolumn{1}{c}{$i$} 
& \multicolumn{1}{c}{$\beta$} 
& \multicolumn{1}{c}{$\gamma$ } 
& \multicolumn{1}{c}{$\gamma-\beta$}
& \multicolumn{1}{c}{$\psi$} 
& \multicolumn{1}{c}{$\alpha(0)$} 
& \multicolumn{1}{c}{$\alpha(\pi)$} 
& \multicolumn{1}{c}{$\cal{D}$} 
\\
\hline 

 2 &  89 &   89.50 &    0.50 &   -2.57 &  -87.50 &   91.50 &    0.00 \\
 6 &  86 &   88.50 &    1.50 &   -2.57 &  -82.50 &   94.50 &    0.00 \\
19 &  80 &   85.01 &    5.01 &   -2.60 &  -66.01 &  104.01 &    0.01 \\
24 &  77 &   83.51 &    6.51 &   -2.62 &  -59.51 &  107.51 &    0.01 \\
29 &  74 &   82.01 &    8.01 &   -2.65 &  -53.01 &  111.01 &    0.02 \\
32 &  72 &   81.01 &    9.01 &   -2.67 &  -49.01 &  113.01 &    0.03 \\
36 &  68 &   79.51 &   10.51 &   -2.71 &  -43.51 &  115.51 &    0.04 \\
39 &  67 &   78.51 &   11.51 &   -2.74 &  -39.51 &  117.51 &    0.04 \\
41 &  65 &   77.51 &   12.51 &   -2.77 &  -36.51 &  118.51 &    0.05 \\
43 &  62 &   76.52 &   13.52 &   -2.80 &  -32.52 &  120.52 &    0.06 \\
46 &  61 &   75.52 &   14.52 &   -2.84 &  -29.52 &  121.52 &    0.07 \\
48 &  59 &   74.52 &   15.52 &   -2.89 &  -26.52 &  122.52 &    0.07 \\
50 &  55 &   73.02 &   17.02 &   -2.96 &  -23.02 &  123.02 &    0.09 \\
51 &  54 &   72.52 &   17.52 &   -2.99 &  -21.52 &  123.52 &    0.09 \\
53 &  53 &   71.53 &   18.53 &   -3.05 &  -18.53 &  124.53 &    0.10 \\
53 &  52 &   71.03 &   19.03 &   -3.08 &  -17.03 &  125.03 &    0.11 \\
55 &  49 &   69.53 &   20.53 &   -3.19 &  -13.53 &  125.53 &    0.13 \\
57 &  48 &   69.03 &   21.03 &   -3.23 &  -12.03 &  126.03 &    0.13 \\
59 &  43 &   67.04 &   23.04 &   -3.40 &   -8.04 &  126.04 &    0.16 \\
60 &  42 &   66.05 &   24.05 &   -3.50 &   -6.05 &  126.05 &    0.17 \\
61 &  41 &   65.55 &   24.55 &   -3.56 &   -4.55 &  126.55 &    0.18 \\
 {\bf 62} & {\bf 37} & {\bf  63.56} & {\bf  26.56} & {\bf  -3.81} & {\bf -0.56} & 
{\bf 126.56} & {\bf   0.20} \\
{\bf 65} & {\bf 33} & {\bf  61.58} & {\bf  28.58} & {\bf  -4.13} & {\bf  3.42} & 
{\bf 126.58} & {\bf   0.23} \\
66 &  31 &   60.59 &   29.59 &   -4.33 &    5.41 &  126.59 &    0.25 \\
67 &  29 &   59.61 &   30.61 &   -4.55 &    7.39 &  126.61 &    0.26 \\
68 &  26 &   58.13 &   32.13 &   -4.95 &    9.87 &  126.13 &    0.29 \\
70 &  20 &   55.23 &   35.23 &   -6.10 &   14.77 &  125.23 &    0.34 \\
71 &  16 &   53.36 &   37.36 &   -7.34 &   17.64 &  124.36 &    0.38 \\
72 &   4 &   51.85 &   47.85 &  -22.91 &   20.15 &  123.85 &    0.62 \\
72 &  14 &   52.47 &   38.47 &   -8.22 &   19.53 &  124.47 &    0.40 \\
73 &   6 &   50.38 &   44.38 &  -16.82 &   22.62 &  123.38 &    0.53 \\
73 &   8 &   50.39 &   42.39 &  -13.26 &   22.61 &  123.39 &    0.48 \\
73 &  10 &   50.91 &   40.91 &  -10.96 &   22.09 &  123.91 &    0.45 \\

\hline
\hline
\end{tabular} 
\end{center} 
\end{table} 

\subsection{A pure magnetic model for the principal pulsation mode} 
\label{sec:models} 
 
Using the formalism of \citet{saio05} we have solved for the effect of a dipole 
magnetic field on the $\nu_1$ mode. In this formalism the mode is assumed to be a 
pure $m = 0$ mode with its axis of symmetry aligned with the magnetic axis; i.e. 
$\gamma = \beta$. We have run a set of models with appropriate $T_{\rm eff}$, 
$\log g$. Unperturbed evolutionary models were calculated by the OPAL opacity with 
a standard chemical composition of $(X,Z)=(0.7,0.02)$. Convective energy flux in 
the envelope is neglected assuming a strong magnetic field suppresses the 
convection. Helium is assumed to be depleted above the first ionization zone, 
treating it in the same way as in \citet{saioetal10}. 

{Here, disregarding the effect of rotation, we assume the pulsations to be 
axisymmetric with respect to the axis of a dipole magnetic field which is inclined 
to the rotation axis by an angle of $\beta$. The eigenfunction of a pulsation mode 
is expressed by a sum of terms proportional to Legendre functions with different 
values of $\ell$. Solving numerically the differential equations for linear non-
adiabatic oscillations including the effects of Lorentz forces, we obtain an 
eigenfrequency and the corresponding eigenfunction; i.e., the relative amplitude 
of each component of the expansion as a function of the depth in the envelope, 
which determines the latitudinal dependence of the pulsation. The latitudinal 
dependence of the luminosity perturbation on the surface predicts the rotational 
modulation of the pulsation amplitude (see \citet{saiogautschy04} and 
\citet{saio05} for details).}

Good fits to the observed amplitude structure seen in Fig.\,\ref{fig:schematic} 
and the amplitude and phase variations as a function of rotation seen in 
Fig.\,\ref{fig:phamp1} were found with a model with $M = 1.70$\,M$_{\odot}$, $\log 
L/{\rm L}_{\odot} = 0.998$, $\log T_{\rm eff} = 3.854$, $\log R/{\rm R}_{\odot} = 
0.314$ and $\log g = 4.04$. Two good matches to the observations were found, one 
with $B_p = 0.7 \pm 0.2$\,kG and the other with $B_p = 15 \pm 2$\,kG. Our high 
resolution spectra clearly rule out a polar field strength as large as 15\,kG, 
since that would produce clear Zeeman splitting, which is not observed, leaving 
only the model with $B_p = 0.7$\,kG. For the latter model, agreement with the 
observed pulsation amplitude and phase variations requires that $\gamma = \beta = 
66^{\circ}$ and $i=62^{\circ}$. 

\begin{figure} 
\centering 
\includegraphics[width=8cm]{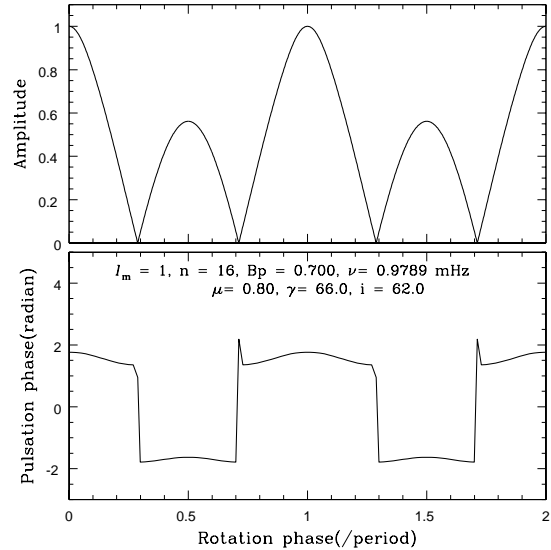} 
\caption{This shows the model pulsation amplitude and phase as a function of 
rotation phase for an $n = 16$ dipole mode with a frequency of $\nu = 
978.9$\,$\mu$Hz, close to the observed $\nu_1 = 974.6$\,$\mu$Hz, and with $i = 
62^{\circ}$, $\gamma = 66^{\circ}$, in good agreement with the oblique pulsator 
model constraints given in Table\,\ref{tab:angles}. } 
\label{fig:modulation} 
\end{figure} 
 
\begin{figure} 
\centering 
\includegraphics[width=8cm]{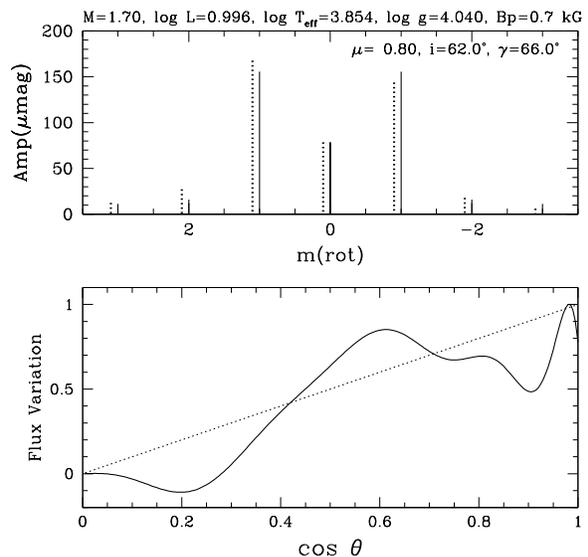} 
\caption{Top: A schematic amplitude spectrum for the model frequencies for 
$\nu_1$. The solid lines show the model amplitudes, while the dotted lines 
(displaced in frequency for visibility) show the observed amplitudes. Bottom: The 
latitudinal dependence of the eigenfunction. The dotted line shows pure dipole 
eigenfunction.} 
\label{fig:multip} 
\end{figure}

Fig.\,\ref{fig:modulation} shows the model amplitude and phase variations as a 
function of rotation phase for an $n = 16$ dipole mode, in good agreement with the 
observations shown in Fig.\,\ref{fig:phamp1}. Fig.\,\ref{fig:multip} compares the 
observed amplitudes of the multiplet structure with the observations. A higher 
than normal value of the limb-darkening coefficient of $\mu = 0.8$ was needed to 
get good agreement for the $A_{\pm 2}$ and $A_{\pm 3}$ components. The 
eigenfunction shown in the bottom panel is significantly deformed from the 
Legendre function $P_1(\cos\theta)$, even though $A_{\pm2}$ and $A_{\pm3}$ are 
much smaller than $A_1$ and $A_0$. This is because for an odd mode such as a 
dipole mode, $l \ge 3$ components (which produce the $A_{\pm2}$ and $A_{\pm3}$ 
sidelobes) are visible only through the limb-darkening effect; without limb-
darkening these components would not be visible due to cancellation on the 
surface. This increased limb-darkening coefficient is consistent with the 
atmospheric temperature gradient in roAp stars, which is steeper than in normal 
stars, as is usually shown by the core-wing anomaly in the Balmer lines 
\citep{cowleyetal01} for most roAp stars. 
 
\subsection{Rotational distortion of the magnetic mode $\nu_1$}
\label{sec:opm}

The magnetic field derived in the previous section is rather weak for roAp stars 
and its effect on the mode inclination can be balanced by the centrifugal force. 
In this case, as shown by \citet{bigot02}, there is no reason to have an axis of 
pulsation aligned along the magnetic axis or the rotation axis. In order to 
calculate the mode inclination $\gamma$ and its polarization $\psi$, we assume a 
value of $\left | \mu^{\rm mag} \right | = 1$ (ratio of magnetic to centrifugal 
shift of frequency). The ratio $\chi$ between the Coriolis and centrifugal shift 
is proportional to $\sim n^{-3}\Omega ^{-1}$, where $n$ is the number of radial 
nodes of the mode. Since $\nu_1$ has a lower frequency and lower $n$ than in 
HR\,3831, the value of $\chi$ is larger than for HR 3831. From the stellar model 
and eigenfunctions used in Section\,\ref{sec:models}, we find $\chi = 0.045$.

Determining the mode inclination from light curves depends strongly on the viewing 
aspect of the mode, i.e. the value of $i$. Indeed, we found several solutions 
($i,\beta$) leading to a fit of the amplitude ratio $(A_{+1}+A_{-1})/A_0$, as long 
as $i< 72^{\circ}$ -- see Table\,\ref{tab:angles}. The values follow roughly the 
law $i +\beta \approx 80-100^{\circ}$. While the angles derived in 
Table\,\ref{tab:angles} lead to the same goodness of fit to the light curve, they 
lead to completely different values of the dipole inclination $\gamma$. Indeed, 
low values of $i$ lead to a dipole almost aligned with the magnetic axis, whereas 
large values lead to a very inclined mode: e.g. 
$\gamma(i=19^{\circ},\beta=80^{\circ}) =85^{\circ} $, 
$\gamma(i=41^{\circ},\beta=65^{\circ}) =77^{\circ} $ and 
$\gamma(i=62^{\circ},\beta=37^{\circ}) = 64^{\circ}$. In all cases, the 
polarization remains small showing that this mode is almost linearly polarized. 
Due to the uncertainty in $i$, it is hard to reach a conclusion about the 
inclination of the principal mode in KIC\,10195926. The same problem appeared in 
the case of HR3831. \citet{bigot02} found a mode well inclined with respect to the 
magnetic axis $\gamma-\beta\approx  26^{\circ}$ based on a value of $i=89^{\circ}$ 
available from spectropolarimetry at this time. A more recent and secure value of 
$i=68^{\circ}$ in HR\,3831 led \citet{bigot10} to a mode almost aligned with the 
magnetic axis $\gamma - \beta\approx 0$. The alignment between magnetic and 
pulsation axes in HR\,3831 seems natural regarding the large magnetic field found 
for the star $\sim 2.5$\,kG. In KIC\,10195926 the lower value of $B_p = 0.7$\,kG 
derived in Section\,\ref{sec:models} suggests a more inclined mode to the magnetic 
field. 

\begin{figure} 
\centering
\includegraphics[width=8cm]{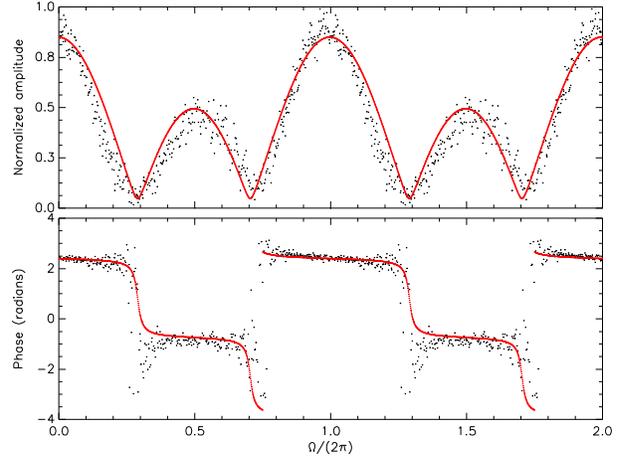} 
\caption{Amplitude (top panel) and phase (bottom panel) of the light curve 
variation corresponding to the mode $\nu_1$ as function of the rotation phase. The 
full line corresponds to a mode tilted to the magnetic axis by $27^{\circ}$ with a 
small polarization ($-4^{\circ}$); $i = 63^{\circ}$. }
\label{fig:opm} 
\end{figure}

We can extract more constraints on the inclination by considering the inequality 
of the side peaks $A_{+1}\neq A_{-1}$. The origin of the inequality is due to the 
Coriolis force which, unlike the magnetic and centrifugal forces, does not act in 
the same way on the $\pm m$ components of the modes. The ratio $(A_{+1}-A_{-
1})/(A_{+1}+A_{-1})$ is therefore very sensitive to the value of $\chi$ which 
determines the polarization of the mode \citep{bigot02}. It also depends on the 
magnetic field and centrifugal forces. Using the constraint on the unequal 
amplitude ratio, we restrict the allowed values of the rotational inclination and 
magnetic obliquity to $i\approx 63^{\circ}$ and $\beta\approx 34^{\circ}$. The 
mode is well inclined to the magnetic axis with $\gamma-\beta \approx 26^{\circ}$. 
We used the formalism developed in \citet{bigot10} to fit of the light curve of 
$\nu_1$. An example is shown in Fig.\,\ref{fig:opm} for ($i=62^{\circ}$, 
$\beta=63^{\circ}$). 

\subsection{The observed amplitude}
\label{sec:amplitude}

The peak pulsation amplitude of KIC\,10195926 occurs at rotation phase zero in 
Fig.\,\ref{fig:lc}. The highest amplitude occurs when the central three 
frequencies of the septuplet are in phase, which gives $(A_{-1}^{(1)} + A_0^{(1)} 
+ A_{+1}^{(1)}) = (168 + 79 + 144)$\,$\mu$mag $\approx 0.4$\,mmag, as is seen in 
Fig.\,\ref{fig:phamp1}. This is potentially detectable in ground-based 
observations, but not easily. However, this is the {\it Kepler} white light 
bandpass amplitude. A study of the roAp star $\alpha$\,Cir using the WIRE 
satellite star-tracker by \citet{brunttetal09} in comparison with simultaneous 
ground-based photometry through a Johnson $B$ filter found a weighted mean 
amplitude ratio of $A_B/A_{\rm WIRE} = 2.3 \pm 0.1$. If we assume that the ratio 
$A_B/A_{\it Kepler}$ is similar, then the maximum amplitude that would be observed 
for KIC\,10195926 in $B$ is expected to be 0.9\,mmag, or peak-to-peak nearly 
2\,mmag. There are many other roAp stars known with similar small amplitudes (see 
Table\,1 of \citealt{kurtzetal06}). 

Columns 6 and 7 of Table\,\ref{tab:angles} give the angle between the line-of-
sight and the pulsation pole for various possible geometries for KIC\,10195926. 
These show that an implausibly high intrinsic amplitude greater than 20\,mmag -- 
more than is observed in any roAp star -- would be needed if $i$ were less than 
$\approx 10^{\circ}$, so those geometries can be ruled out since 
$1/\cos(\alpha(0))$ and $1/\cos(\alpha(\pi))$ are too large ($> 10$).

From the spectroscopically derived values we found in 
Section\,\ref{sec:spectroscopy}  of $T_{\rm eff} = 7200$\,K and $\log g = 3.6$, we 
estimate that $R = 3.6$\,R$_{\odot}$. With the known rotation period $P_{\rm rot} 
= 5.6836$\,d, this radius gives $v_{\rm rot} = 32$\,km\,s$^{-1}$. With the 
measured $v \sin i = 21$\,km\,s$^{-1}$, we then find $i \sim 40^{\circ}$. Although 
there are substantial uncertainties in this estimate, the angles in 
Table\,\ref{tab:angles} suggest that we are seeing nearly the full intrinsic 
amplitude of KIC\,10195926 at rotation phase of best aspect. 

If we use the radius and inclination of our best-fitting model from 
Section\,\ref{sec:models}, $R = 2.1$\,R$_{\odot}$ and $i = 62^{\circ}$, then we 
find that we expect $v \sin i = 17$\,km\,s$^{-1}$. While the errors involved are 
relatively large, this suggests that the model radius and gravity are too low, and 
the spectroscopically derived values from Section\,\ref{sec:spectroscopy} are more 
probable.

\subsection{The harmonic frequencies of $\nu_1$}
\label{sec:harmonic}

\begin{figure} 
\centering 
\includegraphics[width=8cm]{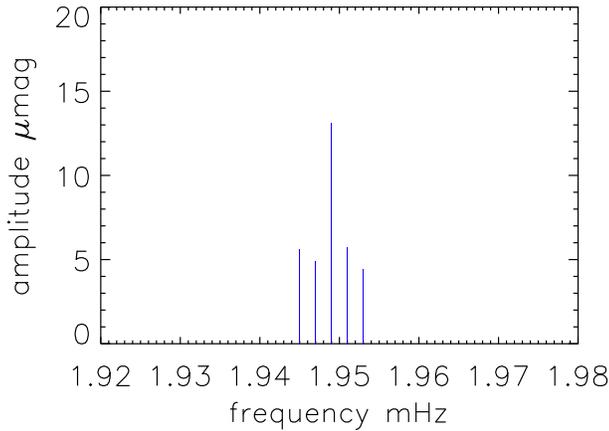} 
\caption{Schematic amplitude spectrum for the $2\nu_1$ harmonic quintuplet.} 
\label{fig:harmonic}
\end{figure}

Fig.\,\ref{fig:harmonic} shows a schematic amplitude spectrum for the quintuplet 
centred on $2\nu_1$. Since the frequency of the central component is twice the 
frequency of the central component of the principal mode septuplet, it appears to 
be the first harmonic of the principal pulsation mode $\nu_1$. Since the principal 
mode $\nu_1$ is also linearly polarized along a fixed axis in the plane defined by 
the magnetic and rotation axes, we can consider that it is an $m=0$ mode with an 
axis inclined by $\gamma$ with respect to rotation axis. 

We therefore use the formalism developed by \citet{kurtzetal90} for nonlinear 
dipole oscillations in roAp stars and replace $\beta$ by $\gamma$. We then expect 
\begin{equation} 
{{A_{+2}^{(2\nu_1)}+A_{-2}^{(2\nu_1)}}\over{A_{+1}^{(2\nu_1)}+A_{-1} 
^{(2\nu_1)}}} ={{1}\over{4}}\tan i \tan \gamma .
\end{equation} 
From the data in Table\,\ref{tab:basicfit} we find for $\nu_1$ that $ 
(A_{+1}^{(\nu_1)} + A_{-1}^{(\nu_1)}) / A_0^{(\nu_1)} = \tan i \tan \gamma = 3.96 
\pm 0.07$. Substituting this value into the above formula leads to 
$(A_{+2}^{(2\nu_1)}+A_{-2}^{(2\nu_1)})/(A_{+1} ^{(2\nu_1)}+A_{-
1}^{(2\nu_1)}))\simeq 1$, which is consistent with the observations. 
 
The amplitude ratios between the central component of the quintuplet of $2\nu_1$ 
and the side components are given by Eqs\,25 and 26 of \citet{kurtzetal90}. If we 
define by $x = \tan i \tan \gamma $ then we have $\left(A_{+1}^{(2\nu_1)}+A_{-
1}^{(2\nu_1)}\right)/A_0^{(2\nu_1)} = x^2/(x^2+2)$ and 
$\left(A_{+2}^{(2\nu_1)}+A_{-2}^{(2\nu_1)}\right)/ A_0^{(2\nu_1)} = 4 x /(x^2+2)$. 
We find for both ratios $\approx 0.89$, which is again fully consistent with the 
observations. 
 
It is reasonable to conclude from these considerations that the quintuplet centred 
on $2\nu_1$ is indeed the first harmonic of the principal pulsation mode $\nu_1$. 
It is potentially possible to derive the values of $i$, $\beta$ and $\mu$ from the 
amplitude ratios of the multiplets of the principal mode $\nu_1$, the secondary 
mode $\nu_2$, and the harmonic $2\nu_1$. 

\subsection{The secondary mode and the frequency spacing}
\label{sec:secondary}

While the frequency septuplet of the primary mode is consistent with our model and 
values of $i$ and $\beta$, the secondary frequency triplet of $\nu_2$ is 
problematic. From the data in Table\,\ref{tab:basicfit} we find for $\nu_1$ that 
\mbox{$(A_{+1}^{(\nu_1)} + A_{-1}^{(\nu_1)}) / A_0^{(\nu_1)} = 3.96 \pm 0.07$}, 
which is significantly different to the value for $\nu_2$ of $(A_{+1}^{(\nu_2)} + 
A_{-1}^{(\nu_2)}) / A_0^{(\nu_2)} = 1.51 \pm 0.23$. This indicates that the two 
modes either arise from different spherical degrees, $l$, or have different 
pulsation axes. 

\begin{figure} 
\centering 
\includegraphics[width=8cm]{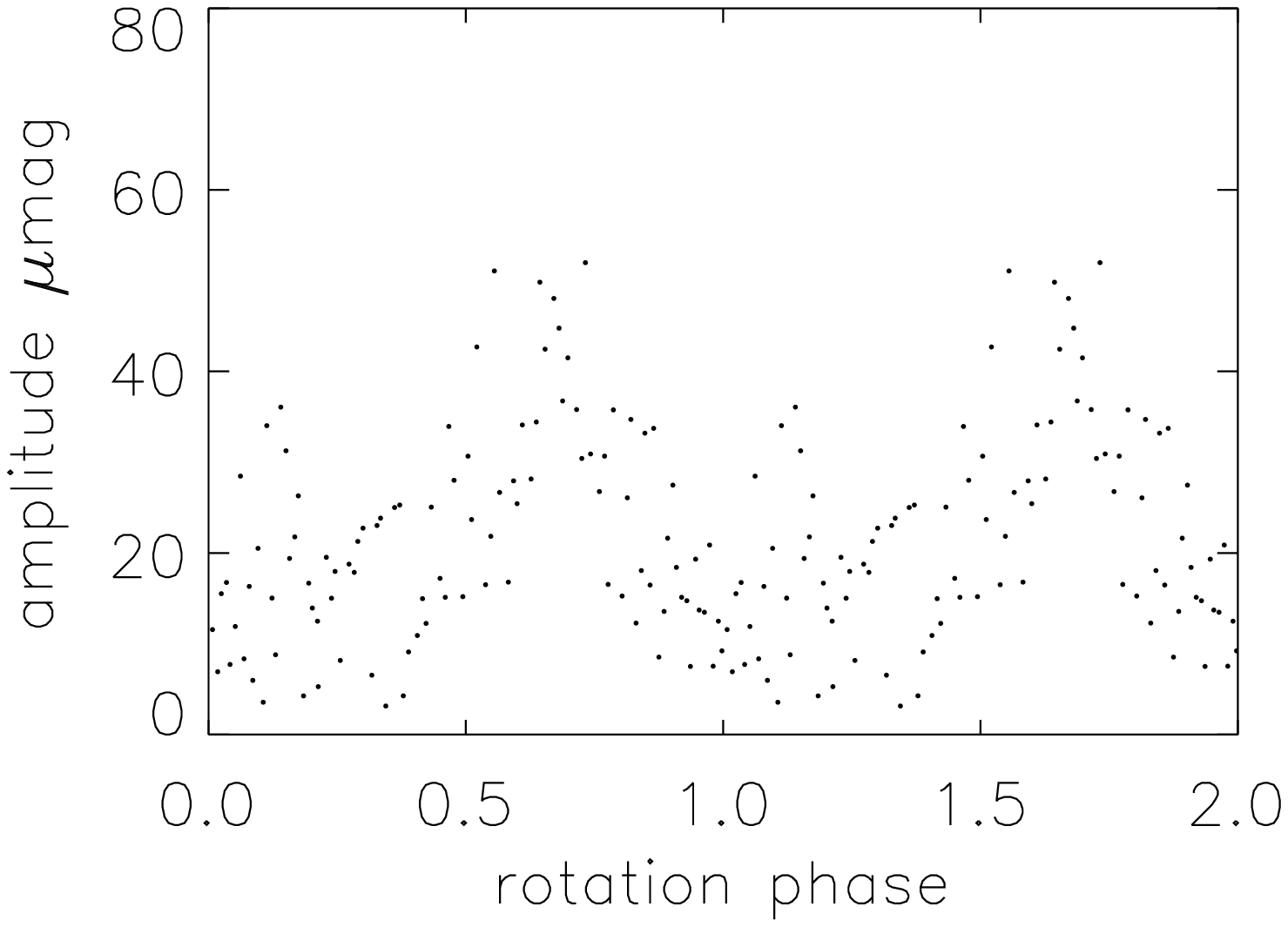} 
\includegraphics[width=8cm]{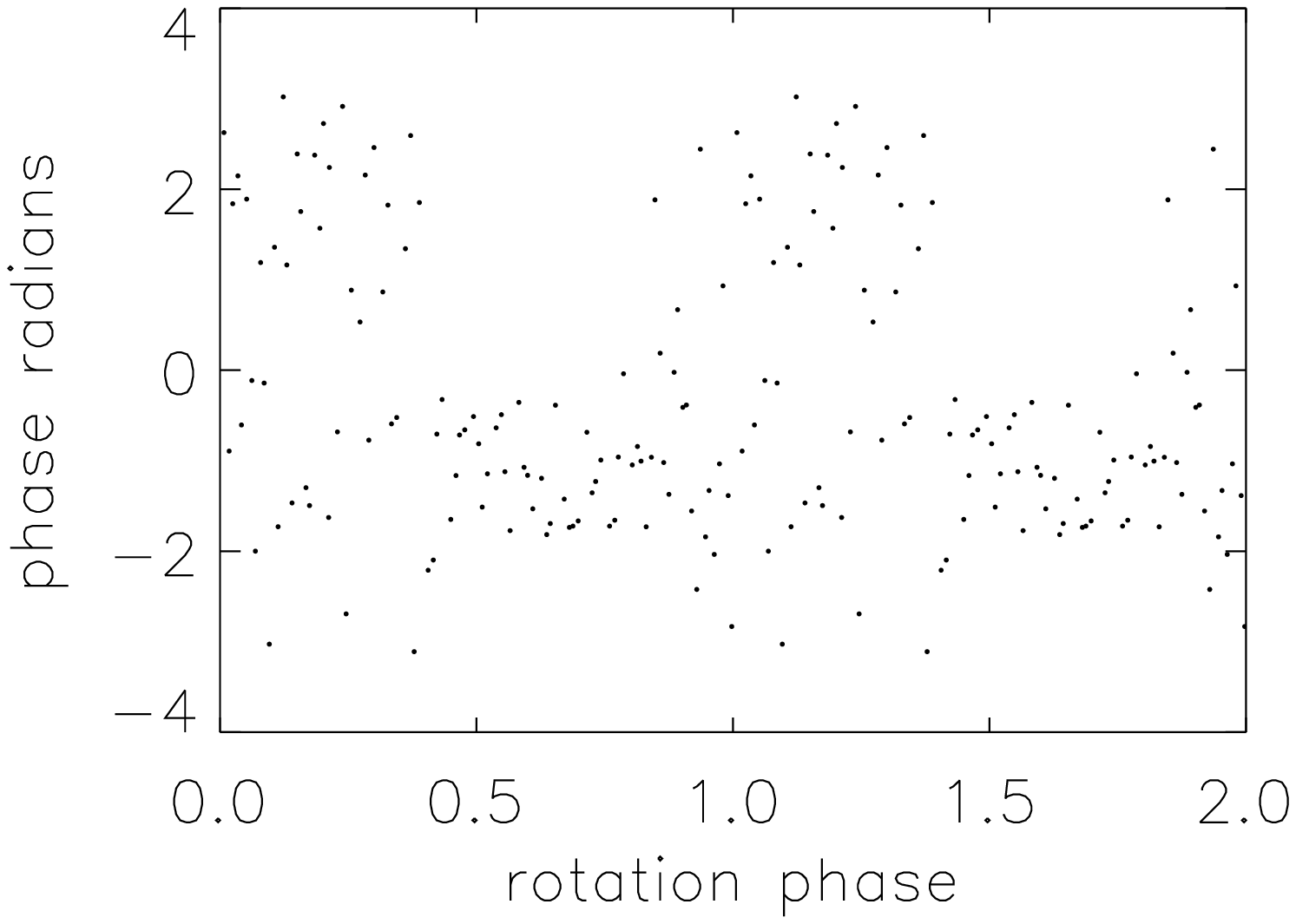} 
\caption{Top: This shows the amplitude modulation as a function of rotational 
phase for $\nu_2$. It has been created by fitting 20 cycles of data by linear 
least squares for $\nu_2$. Notice that the amplitude is modulated only once per 
cycle; both amplitude and phase are poorly defined away from amplitude maximum, 
which occurs around rotation phase 0.67.} 
\label{fig:phamp2}
\end{figure}

We filtered all but the $\nu_2$ triplet from the data, then calculated amplitude 
and phase modulation of that with respect to rotation phase. Because of the 
smaller amplitude, we needed to use 20 cycles for each fit, rather than the 5 we 
used for $\nu_1$. Fig.\,\ref{fig:phamp2} shows the results. Amplitude maximum 
occurs at rotation phase 0.67. There is an appearance of perhaps another maximum 
around rotation phase 0.15; this is probably real, since the rotational sidelobes 
sum to more than the amplitude of the central peak of the triplet, hence two 
maxima per rotation are expected. That the rotational phase of amplitude maximum 
does not coincide with that for $\nu_1$ supports the suggestion that the two modes 
are either of different spherical degree, or have different pulsation axes. We 
also re-ran the least squares fit of Table\,\ref{tab:phasefit}, but shifted the 
zero point in time to get phase equality for the $\nu_2$ triplet, as shown in 
Table\,\ref{tab:nu2fit}. All three frequencies show the same phase within the 
errors for a time of maximum at rotational phase 0.67. This is consistent with 
Fig.\,\ref{fig:phamp2}. 

\begin{table} 
\caption{This table is the same as the part of Table\,\ref{tab:basicfit} 
concerning $\nu_2$, except the zero point in time has been chosen to give phase 
equality for the components of the $\nu_2$ multiplet. This time zero point is 
BJD\,245\,5172.70203; it differs from the time of maximum for $\nu_1$ by 
3.7902\,d, or 0.67 rotation periods. The other frequencies were included in the 
fit, but are not shown here, as only their phases have changed from 
Table\,\ref{tab:basicfit}; they are no longer informative with this different time 
zero point.} 
\label{tab:nu2fit}
\begin{center} 
\begin{tabular}{lrrr} 
\hline 
\multicolumn{1}{c}{ID} 
& \multicolumn{1}{c}{frequency} 
& \multicolumn{1}{c}{amplitude } 
& \multicolumn{1}{c}{phase} \\
\multicolumn{1}{c} {} 
& \multicolumn{1}{c}{$\mu$Hz} 
& \multicolumn{1}{c}{$\mu$mag} 
& \multicolumn{1}{c}{radians} \\
\hline 
$\nu_2 - \nu_{\rm rot}$ &$917.5126$ & $ 8.6 \pm 1.3$ & $ 3.02 \pm 0.15$ \\
$\nu_2$ &$919.5464$ & $ 11.6 \pm 1.3$ & $ 3.00 \pm 0.11$ \\
$\nu_2$ + $\nu_{\rm rot}$ & $921.5801$ & $ 8.6 \pm 1.3$ & $ 3.02 \pm 0.15$ 
\\
\hline
\hline
\end{tabular} 
\end{center} 
\end{table}

The separation of $\nu_1$ and $\nu_2$ is $\nu_1 - \nu_2 = 55.07 \pm 
0.03$\,$\mu$Hz. The model presented in Section\,\ref{sec:models} has a large 
separation $\Delta \nu_0 \approx 57$\,$\mu$Hz, which is close enough to the 
observed separation to suggest that $\nu_2$ is also a dipole mode. If that is 
true, then the two modes do not have the same pulsation axis. On the other hand, 
if we dismiss the model and conclude that the degrees of the two modes are of 
different parity, then 55\,$\mu$Hz cannot be the large separation. 

From $T_{\rm eff} = 7200$\,K and $R = 3.6$\,R$_{\odot}$, we have $\log T_{\rm eff} 
= 3.86$ and $\log L / {\rm L}_{\odot} = 1.5$. From a grid of non-magnetic models 
constructed using the stellar evolution code CESAM \citep{morel97} and 
corresponding pulsation frequencies computed with the code ADIPLS \citep{CD2008b}, 
we find that a model with these global properties has a large separation of 
25\,$\mu$Hz. On the other hand, a large separation of about 2/3 of the observed 
value is realized by a slightly less luminous model, with $\log L / {\rm 
L}_{\odot} = 1.3$, and similar effective temperature. With the case for $\nu_1$ 
being a dipole mode, a plausible alternative is thus that $\nu_2$ is a radial or 
quadrupole mode and that the star is more luminous than what would be expected by 
assuming the observed separation to be the large separation, having, instead, 
$\Delta \nu_0 \sim 37$\,$\mu$Hz. In this case the two observed modes are not 
consecutive in radial overtone; i.e., there is a mode between them that is not 
excited to observable amplitude. This is seen in another roAp star, HD\,217522 
\citep{kreidletal91}, so is not without precedent. The difficulty with this 
solution, when considering the effect of a dipolar magnetic field on the 
oscillations, is that the magnetic distortion of the eigenfunctions of even degree 
modes always generates an $l=0$ component. Even when this component is comparably 
smaller than the $l=2$ component, after averaging over the disk its impact is 
significant. In particular, we were not able to find a magnetic model within the 
scenario of $\nu_2$ being of degree $l=2$, in which the pulsation phase of this 
second mode jumps by an amount comparable to that seen in the observations at 
particular rotation phases.

Fig.\,\ref{fig:phamp2} shows that maximum amplitude for $\nu_2$ is at rotation 
phase 0.67; this is definitely different from the maximum phase of $\nu_1$. Thus 
the rotational phase of maximum and the multiplet amplitude ratios for $\nu_1$ and 
$\nu_2$ are clearly different, yet $\nu_1-\nu_2=55.07\, \mu$Hz is a reasonable 
value as a separation of the two consecutive overtones, suggesting that the modes 
are of the same degree, $l$. A first approach is to treat $\nu_2$ as if it were a 
pure oblique dipole mode perpendicular to $\nu_1$: $\gamma_2 \approx \pi/2 + 
\gamma_1 $.  The maximum amplitude phase $(\sim 0.67)$ and the sub-peak amplitude 
phase $(\sim 0.15)$ coincide with extrema of the rotational phase, thus the 
interpretation of the $\nu_2$ mode as an oblique dipole mode is reasonable. In 
that case, the amplitude ratio is
\begin{equation}
{{\,A_{+1}^{(\nu_2)}+A_{-1}^{(\nu_2)}\,}\over{A_0^{(\nu_2)}}}
\approx{{\tan i} \over{\tan \gamma_1}}.
\end{equation}
However, with the values of the inclinations of the mode $\nu_1$ and the observer 
($i = 62^/circ$, $\gamma_1 = 63.5^\circ$, derived in section 4.3), we cannot fit 
the amplitude ratio of Eq.\,2 within the error bars. We conclude that the dipole 
mode $\nu_2$ is neither aligned nor perpendicular to  $\nu_1$. In Section\,5 a 
tentative explanation is proposed which suggests that the magnetic field acts in a 
different way on the two modes.

\subsection{Asteroseismic position in the HR Diagram} 
\label{sec:HRD} 
 
\begin{figure*} 
\centering 
\includegraphics[width=16cm]{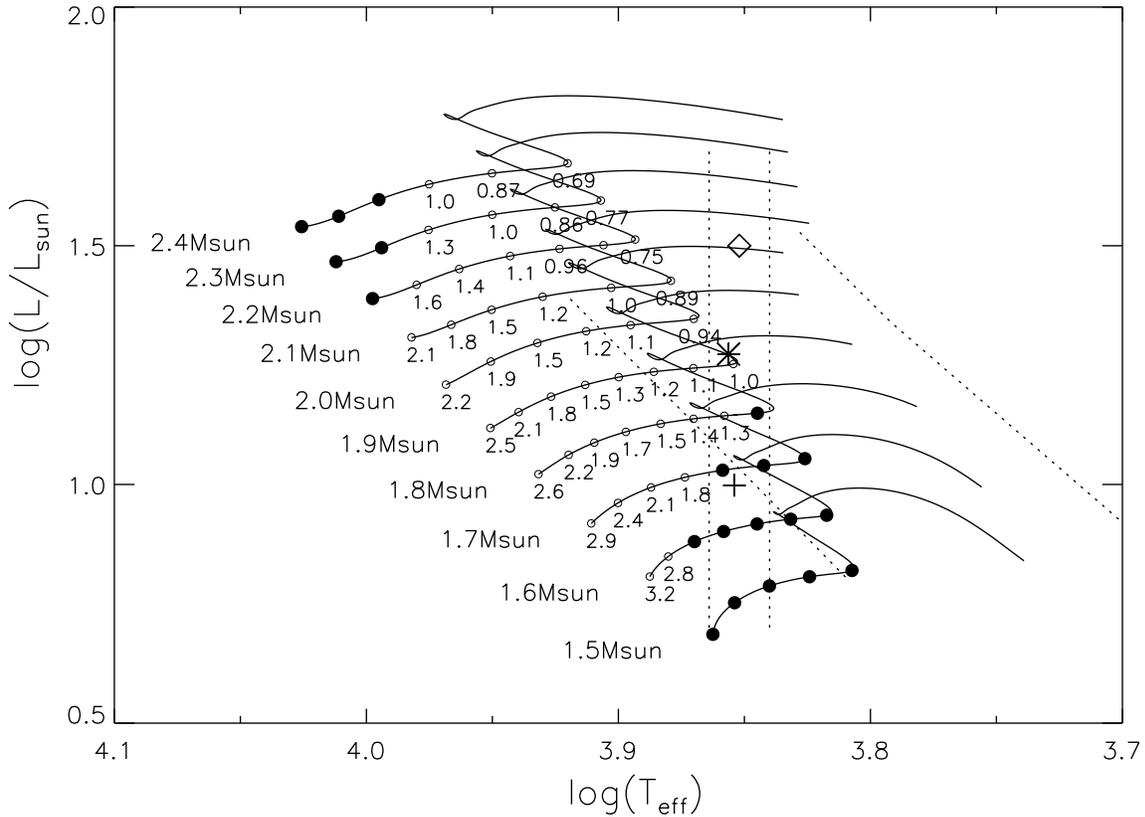} 
\caption{HR diagram with the position of the models discussed in 
Sections\,\ref{sec:models} and \ref{sec:secondary}. The full lines show 
evolutionary tracks for masses between 1.5\,M$_\odot$ and 2.4\,M$_\odot$. The 
numbers along the tracks show the frequency, in mHz, of the most unstable mode for 
each model considered (marked by open circles). The large filled circles show 
models that are stable against pulsations for frequencies in the roAp star range 
(adapted from \citealt{cunha02}). Dashed lines show the ranges of effective 
temperature (vertical) and luminosity (oblique) expected for the star from the 
non-seismic observations alone. The luminosity limits were derived from the grid 
of non-magnetic models, assuming $3.5\le \log g \le 4.0$. The positions of three 
stellar models are shown in this diagram, namely, the magnetic model discussed in 
Section\,\ref{sec:models} (cross), the model discussed in 
Sections\,\ref{sec:amplitude} and \ref{sec:secondary} (diamond) derived from non-
seismic data alone, and the non-magnetic model with large separation of $\sim$ 
37$\mu$Hz, discussed in Section\,\ref{sec:secondary} (star).} 
\label{fig:HRDiagram} 
\end{figure*}

The position of KIC\,10195926 in the Hertzsprung-Russell (HR) Diagram can be 
compared with the theoretical instability strip for roAp stars modelled by 
\citet{cunha02}. The computation of the instability strip, shown in 
Fig.~\ref{fig:HRDiagram}, was based on a pulsational stability analysis of models 
which details are described in \citet{Balmforth2001}. These are non-magnetic 
models, in which the effect of the magnetic field enters only in an indirect way, 
through an assumed suppression of envelope convection. 

The stability against high radial order pulsations of a number of main-sequence 
models of different masses was considered. For each model with unstable high 
radial order modes, the frequency of the mode with largest growth rate was taken 
as a reference for the typical frequencies that may be expected to be observed in 
that region of the HR diagram. The reference frequencies are shown in 
Fig.~\ref{fig:HRDiagram} where we overplotted the three models discussed in the 
previous sections. It is clear that the reference frequencies decrease as the star 
evolves, a phenomenon that is in great part (but not totally) explained by the 
increase in stellar radius \cite[see][for a detailed discussion]{cunha02}.

From Fig.~\ref{fig:HRDiagram} it can be seen that the star is near the terminal 
age main sequence, as expected from its low effective temperature and surface 
gravity. That provides a natural explanation for the low frequency of the modes 
observed in KIC\,10195926, when compared to those seen in the majority of roAp 
stars.

{The models considered in the previous sections, as well as those used to produce 
the instability strip shown in Fig.~\ref{fig:HRDiagram}, all have metallicity 
$Z=0.02$. On the other hand, our analysis of the spectra of KIC\,10195926 in 
Section\,\ref{sec:spectroscopy} indicates that its photospheric mean iron 
abundance is comparable to or slightly higher than the corresponding solar 
abundance of \cite{asplundetal09}. As we pointed out, the global abundances may be 
greater than solar, given the star's young age compared to the Sun and galactic 
chemical evolution. Thus our adopted metallicity in our models of $Z = 0.02$ seems 
a reasonable first choice.} 

{The effects of metallicity on the models need to be explored. This will be done 
when more detailed spectroscopic analysis of the star becomes available. 
Nevertheless, we can anticipate that the main impact of a variation in the 
metallicity adopted in the models will be that of a change of the mass derived for 
the best model of KIC\,10195926. The effect of the metallicity was considered, 
e.g., in the modelling of the roAp star HR\,1217 by \citet{cunhaetal03}, who found 
that for a fixed stellar radius, changing from $Z=0.009$ to $Z=0.019$ led to a 
change of about 4\,$\mu$Hz in $\Delta\nu$ (which is 68\,$\mu$Hz for HR\,1217), as 
a consequence of a change of $\sim$0.2\,M$_\odot$ in the stellar mass. These 
variations, which are also expected if models of different metallicity are 
considered for KIC\,10195926, do not change any of the conclusions presented in 
this work.  We also note that we expect that decreasing the metallicity would 
hardly change the theoretical results presented in Figs.\,11 to 13, although, 
again, the parameters of the best model might be slightly modified.}

\section{Discussion}
\label{sec:discussion}

\subsection{The pulsation axes of the two modes}

KIC\,10195926 shows the power of the precision of the {\it Kepler} photometric 
data. While its principal pulsation at $\nu_1$ could have been discovered in 
ground-based observations, only the $\mu$mag precision of the data has allowed us 
to see examine the rotational multiplet of $\nu_1$, with all the mode geometry 
information that contains. And $\nu_2$ could not be detected from ground-based 
photometry. It is the separation between, and comparison of, $\nu_1$ and $\nu_2$ 
that lead to the remarkable suggestion that the pulsation axes of these two modes 
do not coincide. This is a viable hypothesis for this star supported by the 
improved oblique pulsator model \citep{bigot02}. No other pulsating star has ever 
been noted to have different pulsation axes for different modes. 

In the simple oblique pulsator model \citep{kurtz82} the pulsation axis of the 
star coincides with its magnetic axis. The simple case for the rotational light 
variations of $\alpha^2$\,CVn stars is also for spots that are concentric about 
the magnetic and pulsation poles. It has been clear for decades that magnetic Ap 
stars often have global fields that are more complex than dipoles, and that the 
spot structure is more detailed than two spots at the poles. Recently, 
\citet{brunttetal09} showed that the spots of the roAp star $\alpha$\,Cir do not 
coincide with its pulsation axis. We now see for KIC\,10195926 that $\nu_1$ 
rotational light minimum in brightness coincides with maximum pulsation amplitude, 
whereas for $\nu_2$ maximum pulsation amplitude coincides with one of the 
rotational maxima in brightness. These indicate that the modes are in alignment 
with spots, but not with the same ones, and there is no simple symmetry to the 
spot structure. 

No other roAp star so obviously shows these effects. The number of stars with 
rotational multiplets is small. HD\,6532, HR\,3831 and HR\,1217 are the best 
cases. For the first two there is only one pulsation mode, while the latter shows 
at least 6 modes that have been modelled with distorted modes of different 
spherical degree, but with no thought to use differing pulsation axes 
\citep{saioetal10}. In light of the result here for KIC\,10195926, it will be 
worth revisiting the interpretation of the frequency multiplets for HR\,1217.

The theory of the interaction of rotation, pulsation and magnetic fields is 
complex. \cite{bigot02} presented the improved oblique pulsator model 
incorporating both rotation and magnetic field in the relatively weak field case 
where $B_p \sim 1$\,kG. They showed that the pulsation axis is not necessarily the 
magnetic axis of the star. In their formalism, the inclination of the dipole mode 
depends on how the centrifugal frequency shift compares to the difference between 
magnetic eigenmode frequencies of consecutive azimuthal orders $m$, i.e. 
$\Delta^{\rm mag}_{n,l}=\omega^{\rm mag}_{n,l,m}-\omega^{\rm mag}_{n,l,m+1}$. 

At a given rotation rate, the difference between magnetic and centrifugal shifts 
is a function of the magnetic strength $B_p$ and of the frequency. Therefore, 
strictly speaking two consecutive magnetic overtones of the same degree $l$ must 
have different inclinations with respect to the magnetic field axis. However, as 
shown in \citet{bigotetal00}, the difference $\Delta^{\rm mag}_{n,l} - \Delta^{\rm 
mag}_{n\pm 1,l}$ is very small in most cases, which indicates that consecutive 
modes have roughly the same inclination. Since $\nu_1$ and $\nu_2$ have different 
amplitude ratios we conclude that the two modes have different axes. One 
possibility to explain this is to consider that the two modes are perpendicular. 

Since we found that the $\nu_1$ mode is a dipole almost linearly polarized close 
to the magnetic axis,  a simple interpretation for having a mode at $\nu_2$ with a 
different pulsation axis would have been an orthogonal dipole mode. However, we 
could not find satisfactory agreement in fitting both amplitude ratios 
simultaneously for $\nu_1$ and $\nu_2$, which shows that this hypothesis has to be 
discarded. A mode of degree $l=3$ is also a possibility, but the phase reversal of 
$\nu_2$ occurring between times of pulsation amplitude maxima, as seen in 
Fig.\,\ref{fig:phamp2}, and as deduced from the amplitudes of the $\nu_2$ 
frequency triplet given in Table\,\ref{tab:nu2fit}, suggests the simple geometry 
of a dipole. 

Another possible explanation for the difference in the pulsation axes of the two 
modes observed could come from an effect found by \citet{cunha00} and 
\citet{saiogautschy04}. They clearly showed that at specific frequencies, which 
depend on 
the magnetic field strength, two modes of consecutive radial order are very 
differently affected by the magnetic field. In those cases, the relative size of 
magnetic and centrifugal shifts is likely to be strongly modified from one 
overtone to the next, and the pulsation axes of the two consecutive modes could be 
very different. Further investigation needs to be carried out to test this 
possibility.

\subsection{The rotational subharmonic}

An important issue is the presence of the rotational subharmonic which manifests 
itself as a light variation with twice the rotational period; see 
Fig.\,\ref{fig:lcurvepw}. This is clearly shown by the presence of the subharmonic 
of the orbital frequency at $\frac{1}{2} \nu_{\rm rot}$ in the Fourier Transform 
of independent light curves. We could take the view that this variation is 
external to the star. We could imagine, for example, a companion in an orbit with 
twice the rotational period and exerting a 2:1 resonant tidal effect, in which 
case none of the conclusions so far discussed are affected. However, this seems 
improbable. It will be tested with radial velocity measurements in spectra that we 
are obtaining. 

Let us look for a possible cause of the rotational subharmonics. In what follows 
we note that the theoretical subharmonics appear at $\nu = \nu_{\rm rot}/3$, 
$2\nu_{\rm rot}/5$, $\nu_{\rm rot}/2$, $2\nu_{\rm rot}/3$, $3\nu_{\rm rot}/4$, 
$\ldots$, as well as $4\nu_{\rm rot}/3$, $3\nu_{\rm rot}/2$ and $2\nu_{\rm rot}$. 
These are a systematic sequence. The observed subharmonics are clearly seen at 
$\nu_{\rm rot}/2$ and $3\nu_{\rm rot}/2$. However, least-squares fitting of other 
possible subharmonics shows that some of them have formally significant 
amplitudes. Because of the low frequency noise in the data caused by instrumental 
drift -- particularly in Q1 -- and because a gap in the data between Q1 and Q3.3 
introduces aliases, we are not confident in the formal errors. In what follows the 
frequencies proposed are consistent with the data, with $\nu_{\rm rot}/2$ and 
$3\nu_{\rm rot}/2$ having the highest confidence. 

Our hypothesis is that torsional modes, or r modes, are excited and generate the 
observed subharmonic frequencies. In a rotating star, the radial component of 
vorticity interacts with the Coriolis force and disturbs the equilibrium structure 
of the star. This generates r-mode oscillations, which are torsional oscillations 
(Unno et al. 1989).  
 
As the light curve varies with the rotation period, there are brightness 
inhomogeneities on the stellar surface, which may be decomposed into a series of 
spherical harmonics. The mean flux distribution averaged over latitude  is then 
expressed in the corotating frame as 
\begin{equation} 
F \propto  
\left(1+\sum_{m'\neq 0}\alpha_{m'} 
\exp(im'\phi_{\rm R})\right), 
\end{equation} 
where $\alpha_{m'}$ denotes the amplitude of brightness inhomogeneity of  $m'$-
component, and $\phi_{\rm R}$ is the azimuthal angle in the corotating frame.  
 
By treating such a patchy configuration as the equilibrium state and considering 
perturbations on it, we expect the flux variation caused by the r mode with the 
angular degree $l$ and the azimuthal order $m$ is expressed as 
\begin{equation} 
{{\delta F}\over{F}} 
\propto 
\exp(im\phi_{\rm R})\exp(i\nu_{\rm r}t), 
\end{equation} 
where $\nu_{\rm r}$ denotes the frequency of the r mode  with the degree $l$ and 
the azimuthal order $m$ in the corotating frame  and it is given, to the first 
order, as 
\begin{equation} 
\nu_{\rm r} ={{2m\nu_{\rm rot}}\over{l(l + 1)}}.  
\end{equation} 
If $m = l$, $\nu_{\rm r}$ is reduced to $2\nu_{\rm rot}/(l + 1)$:  
$2\nu_{\rm rot}/3$ for $l = 2$;  $\nu_{\rm rot}/2$ for $l = 3$; 
$2\nu_{\rm rot}/5$ for $l = 4$, $\ldots$
 
In order to move from the corotating frame to an inertial frame, we only have to 
utilize the following transformation;  $\phi_{\rm R}=\phi_{\rm I}-\nu_{\rm rot}t$, 
where $\phi_{\rm I}$ is the azimuthal angle in the inertial frame.   The 
observable luminosity variation is then given by 
\begin{eqnarray}
\delta L
&\propto&
\exp[im\phi_{\rm I}+i2\pi(\nu_{\rm r}-m\nu_{\rm rot})t\} 
\nonumber\\
& & 
+\sum_{m'\neq 0}
\left[
\alpha_{m'}\exp\{i(m+m')\phi_{\rm I}
\right.
\nonumber\\
& &
\left.
+i2\pi\{\nu_{\rm r} - (m+m')\nu_{\rm rot} \}t]
\right].
\end{eqnarray}
The component of $m'=-m$ leads the corotating-frame frequency of the r mode,  
$\nu_{\rm r}$, to be observable.  The visibility of the corotating-frame 
frequencies depends on the size and the contrast of the surface brightness 
inhomogeneity. 
  
Note that both the corotating-frame and the inertial-frame frequencies are 
observable. As an example, we fitted by least-squares $4\nu_{\rm rot}/3 = 
0.234552$\,d$^{-1}$ to the Q3.3 data and found that it has a highly significant 
amplitude of $52 \pm 2$\,$\mu$mag. In our interpretation this is a manifestation 
of the inertial-frame frequencies, $m\nu_{\rm rot}[1-2/l(l+1)]$ for $l=m=2$. These 
considerations suggest that our new finding of the subharmonics is essentially 
discovery of torsional modes. New data are being acquired by {\it Kepler} for this 
star, with which we will test this idea further. 

\section{Future work}

{\it Kepler} is continuing to observe KIC\,10195926 in short cadence, so we will 
have a longer data set with higher S/N to work with in the future. We have 
obtained high resolution spectra and will gather more in future observing seasons. 
These will provide better constraints on $T_{\rm  eff}$ and $\log g$, which 
because of their current errors leave uncertainties in the interpretation of 
$\nu_1$ and $\nu_2$ within our models, hence uncertainty in the large separation. 
KIC\,10195926 is clearly a spectrum variable, and we plan Doppler Imaging studies 
of its spectral line variations to map surface abundance distributions; we also 
plan model reconstruction of the spots that lead to the rotational photometric 
variations. Spectropolarimetry is planned to determine the magnetic field strength 
and structure. KIC\,10195926 is an outstanding example of the power of the time 
span, duty cycle and precision of the {\it Kepler} data for asteroseismology, and 
it demonstrates well the need and usefulness of ground-based follow-up 
observations using many medium aperture telescopes. 

\section{acknowledgements}
Funding for the {\it Kepler} Mission is provided by NASA's Science Mission 
Directorate. The authors gratefully acknowledge the {\it Kepler} Science Team and 
all those who have contributed to making the {\it Kepler} Mission possible. Some 
data are based on observations made with the Nordic Optical Telescope, operated on 
the island of La Palma jointly by Denmark, Finland, Iceland, Norway, and Sweden, 
in the Spanish Observatorio del Roque de los Muchachos of the Instituto de 
Astrofisica de Canarias. This work was partially supported by the project 
PTDC/CTE-AST/098754/2008 and the grant SFRH /BD / 41213 / 2007 funded by 
FCT/MCTES, Portugal. MC is supported by a Ciencia 2007 contract, funded by 
FCT/MCTES(Portugal) and POPH/FSE (EC). MG received financial support from an NSERC 
Vanier scholarship. DWK and VGE are supported by the Science and Technology 
Facilities Council of the UK. 

\bibliography{revised_10195926} 
\end{document}